\documentclass[aps,prl,reprint,twocolumn,floatfix,superscriptaddress,longbibliography]{revtex4-2}

\usepackage{standalone}
\usepackage{amsthm}
\usepackage{lipsum}
\usepackage{amsfonts, amsmath, amssymb}
\usepackage{graphicx}
\usepackage{dcolumn}
\usepackage{bm}
\usepackage{dsfont}
\usepackage[normalem]{ulem}
\usepackage{color}
\usepackage[utf8]{inputenc}
\usepackage[colorlinks,citecolor=blue,linkcolor=blue,urlcolor=blue]{hyperref}
\usepackage{tikz}
\usepackage{braket}
\usepackage{physics}
\usepackage{color}
\usepackage{soul}
\usepackage{mathtools}
\usepackage{float}
\usepackage{esvect}
\usepackage{subfigure}
\usepackage{stmaryrd} 
\usepackage{enumerate}
\usepackage{bbold}

\include{physics}

\begin{document}

\title{Electron-photon Chern number in cavity-embedded 2D moir\'e materials}
\author{Danh-Phuong Nguyen}
\affiliation{Universit\'{e} Paris Cit\'e, CNRS, Mat\'{e}riaux et Ph\'{e}nom\`{e}nes Quantiques, 75013 Paris, France}
\author{Geva Arwas}
\affiliation{Universit\'{e} Paris Cit\'e, CNRS, Mat\'{e}riaux et Ph\'{e}nom\`{e}nes Quantiques, 75013 Paris, France}
\author{Zuzhang Lin}
\affiliation{Department of Physics, The University of Hong Kong, Hong Kong, China}
\affiliation{HKU-UCAS Joint Institute of Theoretical and Computational Physics at Hong Kong, China} 
\author{Wang Yao}
\affiliation{Department of Physics, The University of Hong Kong, Hong Kong, China}
\affiliation{HKU-UCAS Joint Institute of Theoretical and Computational Physics at Hong Kong, China}
\author{ Cristiano Ciuti}
\affiliation{Universit\'{e} Paris Cit\'e, CNRS, Mat\'{e}riaux et Ph\'{e}nom\`{e}nes Quantiques, 75013 Paris, France}

\begin{abstract}
\noindent 

We explore theoretically how the topological properties of 2D materials can be manipulated by cavity quantum electromagnetic fields for both resonant and off-resonant electron-photon coupling, with a focus on van der Waals moir\'{e} superlattices. We investigate an electron-photon topological Chern number for the cavity-dressed energy minibands that is well defined for any degree of hy- bridization of the electron and photon states. While an off-resonant cavity mode can renormalize electronic topological phases that exist without cavity coupling, we show that when the cavity mode is resonant to electronic miniband transitions, new and higher electron-photon Chern numbers can emerge.

\end{abstract}
\date{\today}
\maketitle
In recent years van der Waals heterostructures combining atomic-thin layers of 2D materials such as graphene or transition metal dichalcogenides (TMD) have attracted a great deal of interest \cite {Geim2013,Ajayan2016,Novoselov2016,Song2018,valosOvando2019,Yankowitz2019}. Indeed, these systems present rich and controllable physical properties already at the single-particle level due to multiple quantum degrees of freedom, namely the electron spin, the valley and layer pseudospins. This class of 2D materials exhibits a wide variety of interesting electronic properties including semimetallic, semiconducting, superconducting and magnetic phases. One prominent example is magic angle twisted bilayer graphene, exhibiting quasi-flat electronic bands \cite{Bistritzer2011} and remarkable superconducting properties \cite{Cao2018}. Another notable class of moir\'{e} heterostructures is based on TMD materials, which, thanks to their semiconductor properties, have particularly interesting optical properties \cite{Seyler2019,Tran2019,Jin2019}.

A growing interest is emerging for the manipulation of materials with cavity vacuum fields \cite{FornDaz2019, FriskKockum2019,GarciaVidal2021,Schlawin2022}. In particular, metallic split-ring terahertz electromagnetic resonators are remarkable for their deeply sub-wavelength photon mode confinement \cite{Keller2017,Scalari2012, ParaviciniBagliani2018} with mode volumes that can be as small as $10^{-10}\lambda_0^3$, being $\lambda_0$ the free-space electromagnetic wavelength corresponding to the resonator frequency.  
Studies on GaAs 2D electron gases have shown that electronic transport in mesoscopic quantum Hall bars can be greatly modified by the coupling to electromagnetic resonators even without illumination \cite{Appugliese2022}, as a result of  cavity-mediated electron hopping \cite{Ciuti2021,Arwas2023}, which results in a breakdown of the Hall resistance quantization associated to the topological properties of the electronic Landau states.

Recent theoretical works have investigated how to exploit cavity QED to control topological properties of systems, such as 1D tight-binding chains described by the SSH model \cite{Dmytruk2022}. Concerning 2D systems, a recent work has studied 2D bulk materials \cite{Wang2019,Li2022} where the standard electron Chern number is computed by considering an effective electronic Hamiltonian obtained by adiabatic elimination of the photon degrees of freedom. A letter exploring single-sheet graphene ribbons \cite{Guerci_PRL_2020} has studied  electron Chern numbers computed once the cavity field is approximated in a classical coherent state. Another investigation has explored a generic single-electron problem in the ultrastrong light-matter coupling regime \cite{Masuki2023} and focused on the topological control in the configuration where the cavity photon mode frequency is much larger than the relevant electronic transition frequencies. A key point that has not been addressed is the behavior of genuine electron-photon topological invariants that are associated to the interacting quantum electron-photon system when the photon mode is resonant to electronic transitions.  In the past, this has been done only for classical exciton-polariton normal modes where a bosonic exciton field is strongly coupled to a cavity photon field \cite{Karzig_PRX_2015,Ozawa_RMP_2019} (the bare cavity photon and exciton bands are topologically trivial, but the hybrid light-matter polariton bands      are not). However, for a fermionic particle coupled to a quantized cavity field, to the best of our knowledge, the study of  electron-photon topological invariants for the resonant light-matter coupling has been overlooked.

In this letter, we explore the properties of electron-photon Chern numbers, focusing on cavity-embedded 2D moir\'e materials. We investigate different regimes of coupling (off-resonant versus resonant cavity mode), for different cavity geometries (mode with in-plane or out-of-plane polarization) and explore the topological transitions characterized by such an electron-photon Chern number with realistic values for state-of-the-art split-ring resonators and TMD materials. We show that in the case of resonant electron-photon coupling, new topological phases and high Chern numbers can emerge.

\begin{figure}
\centering
\includegraphics[width = 0.95\hsize]{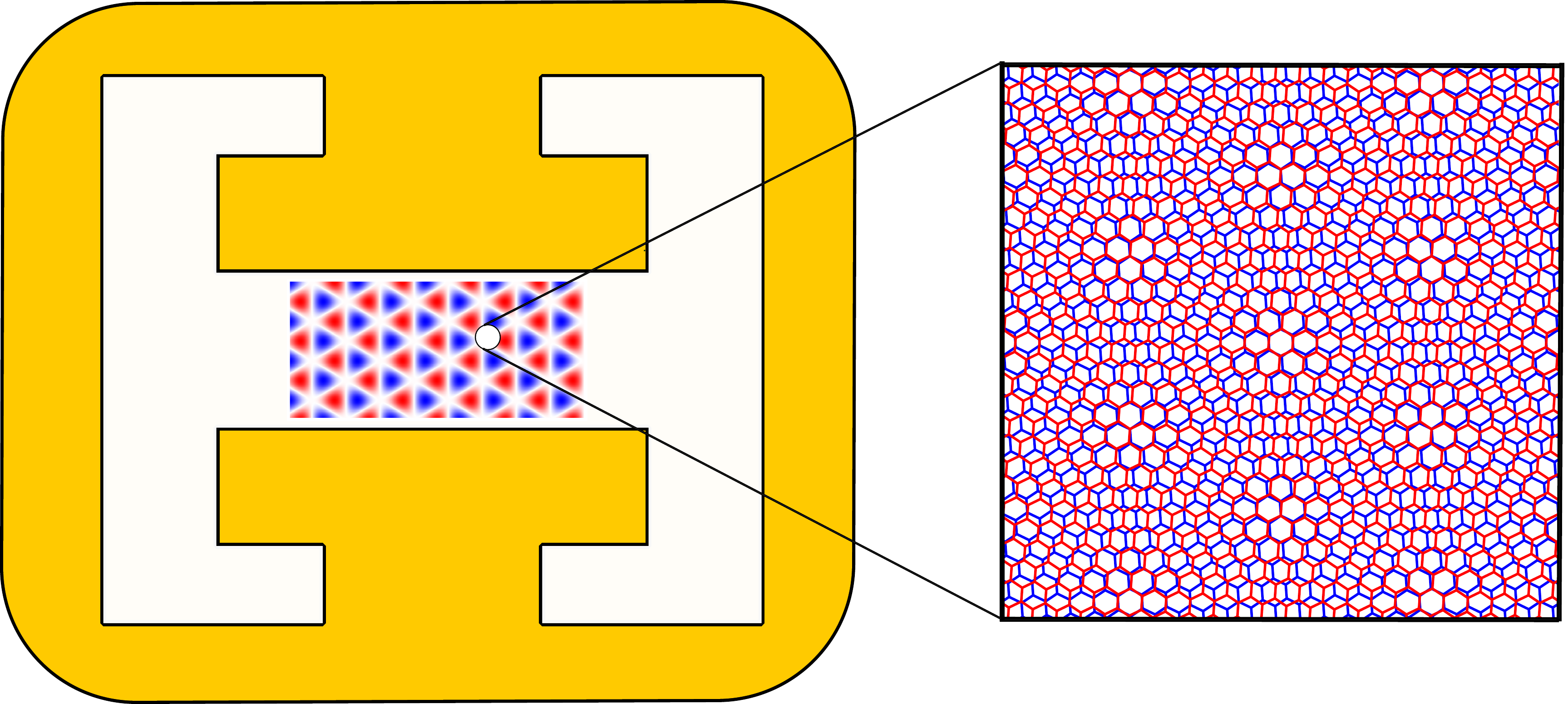} \\ 
\includegraphics[width = 0.49\hsize]{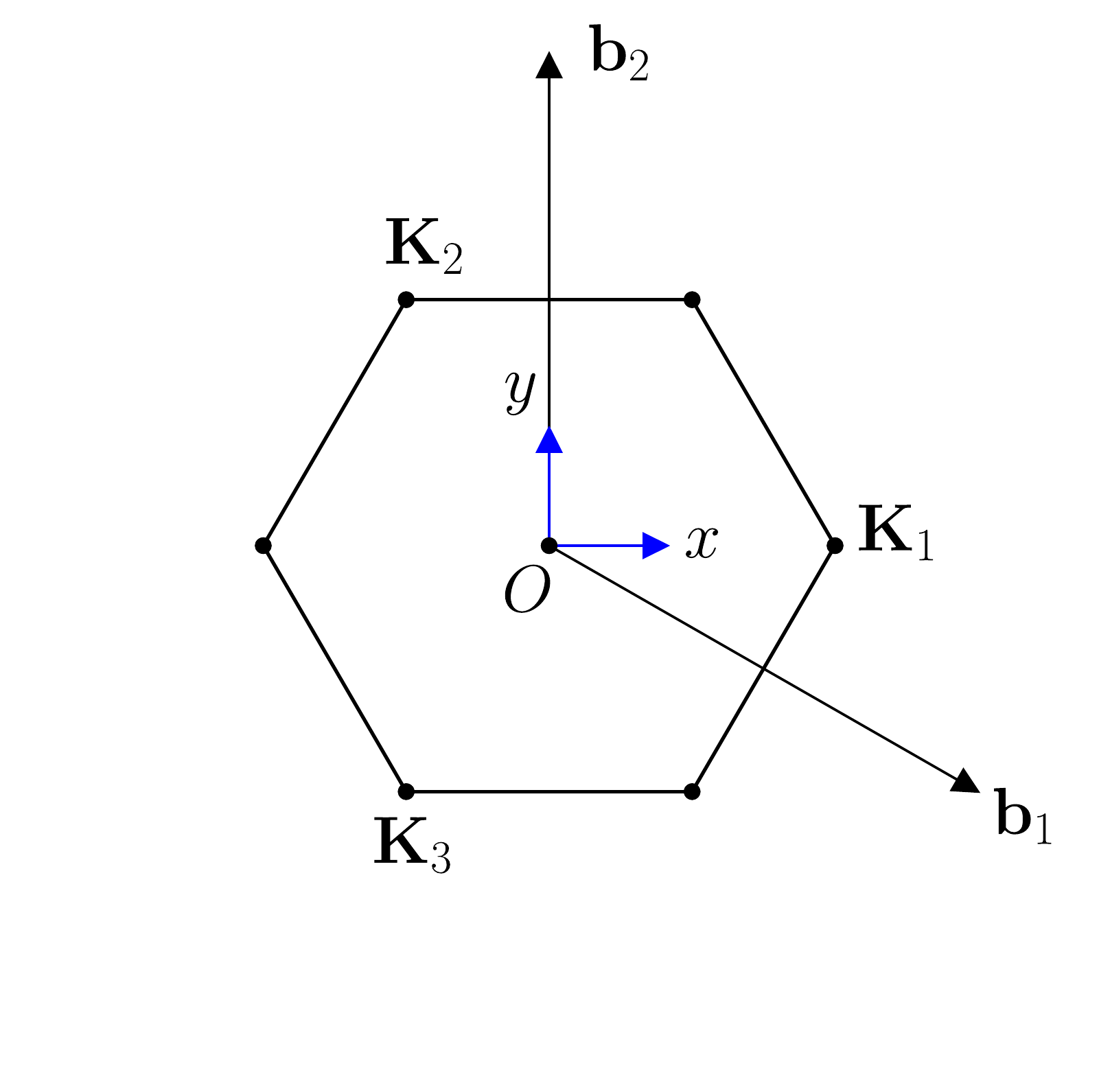}
\includegraphics[width = 0.49\hsize]{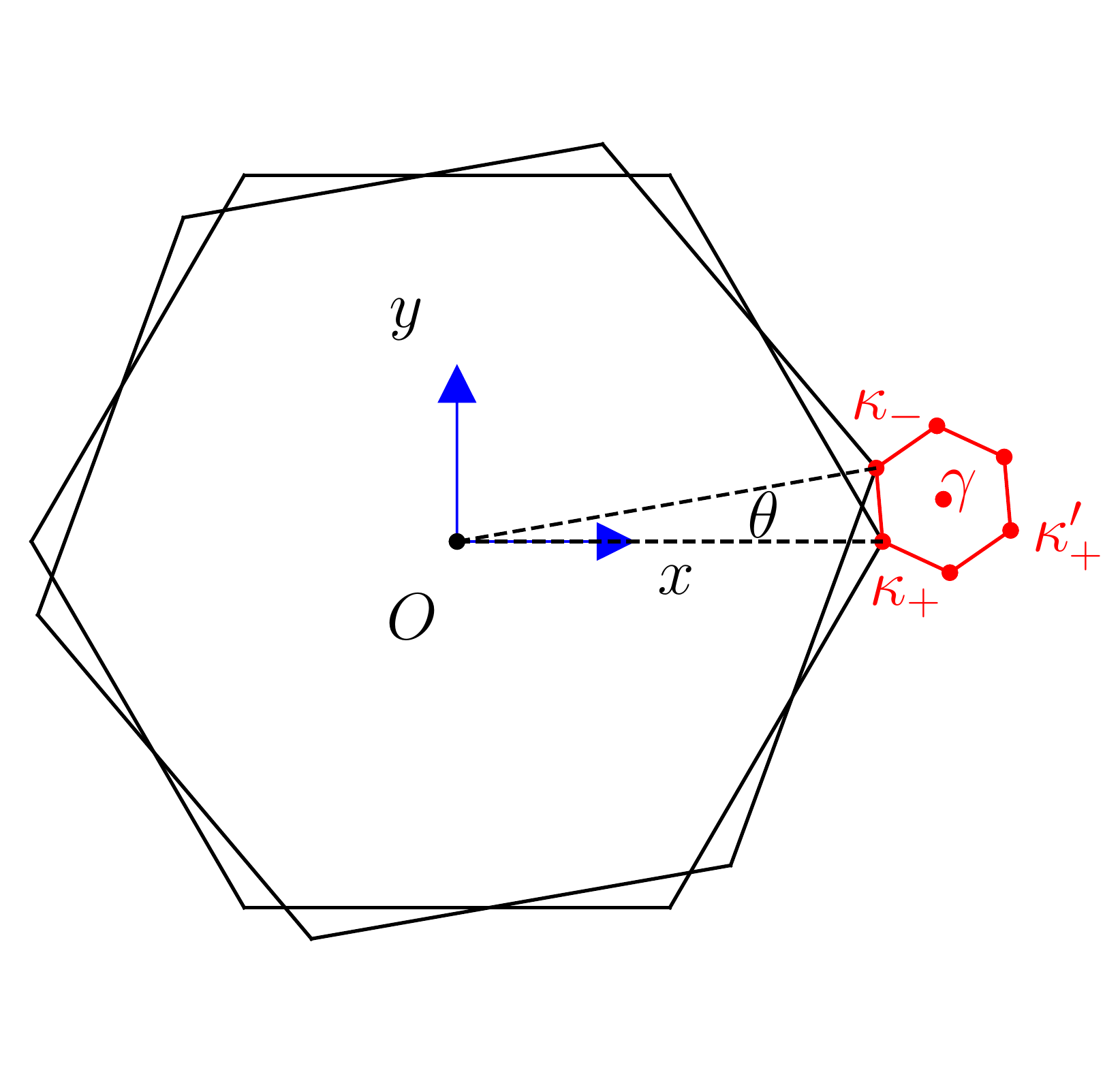}
\caption{Top panel: sketch of a twisted bilayer system with its moir\'e pattern inside the gap region of a split-ring resonator. Bottom panel: on the left, the first Brillouin zone with Dirac points of the bottom layer and corresponding wavevectors; on the right, the Brillouin zones of both the bottom and rotated top layer together with the moir\'{e} mini Brillouin zone (shown in red) for the corresponding moir\'e superlattice.}\label{fig:TMD-mBZ}
 
\end{figure}

{\it Cavity QED Hamiltonian ---} Let us consider a bilayer moir\'e system consisting of two twisted TMD layers embedded in a single-mode resonator (see the sketch in Fig. \ref{fig:TMD-mBZ}). Each of them is a honeycomb lattice defined by two primitive translation vectors $\mathbf{a}_1 = a_0\sqrt{3}/2(1, \sqrt{3}, 0)$ and $\mathbf{a}_2 = a_0\sqrt{3}/2(-1, \sqrt{3}, 0)$, where $a_0$ is the monolayer TMD lattice constant. We consider in-plane parallel stacking (R-stacking) with the layer on top rotated by a small angle $\theta$ in order to create a long wavelength moir\'{e} pattern. The distance between two layers is $d$. The moir\'{e} unit cell is defined by $\mathbf{L}_{i = 1,2} = \left[\mathbb{1} - R(\theta)^{-1}\right]^{-1}\mathbf{a}_i$, where $\mathbb{1}$ and $R(\theta)$ are respectively the identity and the rotation matrix corresponding to the rotation angle $\theta$ about the $z$-axis. We denote the moir\'{e} lattice constant $\left|\mathbf{L}_i\right|$ as $a_M$, where for small angles $a_{M} \simeq a_0/\theta$. In the following, for numerical applications we will use parameters of the $\text{MoTe}_2$ system \cite{Wu2019}. For the photonic part, thanks to the approximately flat mode inside the capacitive gap of complementary split-ring resonators \cite{Appugliese2022}, we will consider a single-mode cavity with a spatially homogeneous field described by the vector potential $\hat{\mathbf{A}} = A_0 \, \mathbf{u}(\hat{a} +\hat{a}^{\dagger})$ with $\hat{a}^{\dagger}$ ($\hat{a}$) the photonic creation (annihilation) operator. The vacuum field amplitude is $A_{\rm 0} = \sqrt{\hbar/(2\omega_c\epsilon_0V_{mode}})$ with $\omega_c$ the mode frequency, $V_{mode} = \eta\lambda_c^3$ the effective mode volume related to the compression factor $\eta$ and $\lambda_c$ the wavelength in free space that corresponds to $\omega_c$. The orientation of the cavity mode polarization is described by the unit vector $\mathbf{u} = (u_x, u_y, u_z)$.

Our treatment is based on the four-band continuum model for small angle twisted bilayer TMD \cite{Wu2019,Zhai2020}. The cavity QED Hamiltonian can be written as
\begin{equation}
\label{eq:Haldane-H4}
\hat{H} = \hbar\omega_c\hat{a}^{\dagger}\hat{a} + \begin{pmatrix}
        \hat{H}_{t} & 0 \\ 0 & \hat{H}_{b}
    \end{pmatrix} + \begin{pmatrix}
        \hat{V}_t & \hat{U} \\ \hat{U}^{\dagger} & \hat{V}_b
    \end{pmatrix},
\end{equation}
where the $2\cross2$ operator matrix $\hat{H}_{t}$ ($\hat{H}_{b}$)  corresponds to the conduction and valence bands of the top (bottom) layer. The moir\'{e} potentials are given by the last term.
The light-matter coupling is introduced via Peierls factors \footnote{See Supplementary Material for additional details about the cavity QED Hamiltonian, the purity of electronic reduced density matrix, the spectral functions and spatial distribution of edge states.} in the intralayer ($\hat{H}_{t}$, $\hat{H}_{b}$) and interlayer ($\hat{U}$) hopping terms.
We quantify the interaction strength by the dimensionless constant $g = eA_0 a_0/\hbar$.
For $g \ll 1$ and in the low-energy sector, each diagonal block matrix $\nu = t, b$ and interlayer coupling $\hat{U}$ can be approximated as follows:
\begin{equation}
\label{eq:TMD-each H}
\begin{aligned}
    &\hat{H}_{\nu} = \Delta \sigma_z + v_F\boldsymbol{\sigma}_{xy} \cdot \left[\mathbf{\hat{p}} -\hbar\mathbf{K}_{\nu} + e\mathbf{A}^{(xy)}(\hat{a}+\hat{a}^{\dagger})\right], \\
    &\hat{U} = e^{-i\frac{eA^{(z)}d}{\hbar}(\hat{a}+\hat{a}^{\dagger})}\hat{U}_0,
    \end{aligned}
\end{equation}
with $\mathbf{K}_{t} = \boldsymbol{\kappa}_-$, $ \mathbf{K}_{b} = \boldsymbol{\kappa}_+$ shown in Fig. \ref{fig:TMD-mBZ}, $\boldsymbol{\sigma}_{xy} = (\sigma_x, \sigma_y, 0)$ a vector of Pauli matrices, $\mathbf{A}^{(xy)}$ is the in-plane projection of $A_0 {\mathbf u}$ and $A^{(z)}$ is its $z$-component. Note that the interlayer distance $d$ is much  larger than the in-plane lattice constant $a_0$. Hence, the argument of the Peierl's  exponential for the $U$ term is much larger than its $t$ counterpart (in-plane coupling). We will consider a regime of realistic parameters, where it turns out that the exponential of the $t$-term can be linearized, while the out-of-plane Peierls term must be kept in exponential form. The moir\'{e} potentials $\hat{V}_{\nu}$ and $\hat{U}_0$ are the same as the ones in \cite{Wu2019,Zhai2020}. Due to the large energy gap between the conduction and valence band in the TMD material, we can restrict our description to the latter, for which non-trivial topological properties were studied \cite{Wu2019,yu_giant_2020} in the absence of a cavity field. Finally, we represent the Hamiltonian in the hole picture. The corresponding Hamiltonian reads:
\begin{widetext}
\begin{equation}
\label{eq:TMD-H1}
\begin{aligned}
    \hat{\mathcal{H}} = \hbar \omega_c \hat{a}^{\dagger}\hat{a}
    +\frac{1}{2m^{\star}}\begin{pmatrix}
   (\hat{\mathbf{p}} + \hbar\boldsymbol{\kappa}_-  - e\mathbf{A}^{(xy)}(\hat{a}+\hat{a}^{\dagger}))^2 & 0 \\
    0 & (\hat{\mathbf{p}} + \hbar\boldsymbol{\kappa}_+ -e\mathbf{A}^{(xy)}(\hat{a}+\hat{a}^{\dagger}))^2
    \end{pmatrix} 
    -  
    \begin{pmatrix}
    \hat{V}^v_t & \hat{U}_0^{v\dagger} \\ \hat{U}_0^{v} & \hat{V}^v_b    \end{pmatrix} - i\frac{\omega_ceA^{(z)}d}{2}(\hat{a}-\hat{a}^{\dagger})\tau_z
\end{aligned}
\end{equation}
\end{widetext}
where $m^{\star} = \Delta/v_F^2$ is the effective mass of the valence band, $\tau_z$ is the $z$-axis Pauli matrix with respect to the layer pseudospin. We have considered minimal coupling in the Coulomb gauge for the in-plane motion, while the dipole gauge resulting from a Power-Zienau-Woolley (PZW) transformation is used for the out-of-plane part. The expression for the moir\'{e} potentials $\hat{V}_t^v$, $\hat{V}_b^v$, $\hat{U}_0^v$ is the same as in \cite{Wu2019,Zhai2020} and is reported in the Supplemental Material. To account for the interlayer bias $V_z$ (that breaks the symmetry between $\boldsymbol{\kappa}_{+}$ and $\boldsymbol{\kappa}_-$ \cite{Wu2019}), the term $-V_z/2 \cross \tau_z$ has to be added to the Hamiltonian in Eq.  (\ref{eq:TMD-H1}). As the moir\'{e} potential is periodic with respect to $\mathbf{L}_{i=1,2}$ translations, the electronic part can be block-diagonalized in momentum space (Bloch theorem). Each block $\mathbf{k}$ belongs to the moir\'{e} mini Brillouin zone (mBZ) shown in Fig. \ref{fig:TMD-mBZ}, spanned by moir\'{e} reciprocal vectors $\mathbf{G}_1$ and $\mathbf{G}_2$ satisfying $\mathbf{G}_i\cdot\mathbf{L}_j = 2\pi\delta_{ij}$. 
\begin{figure}[t!]
\includegraphics[width = 0.95\hsize]{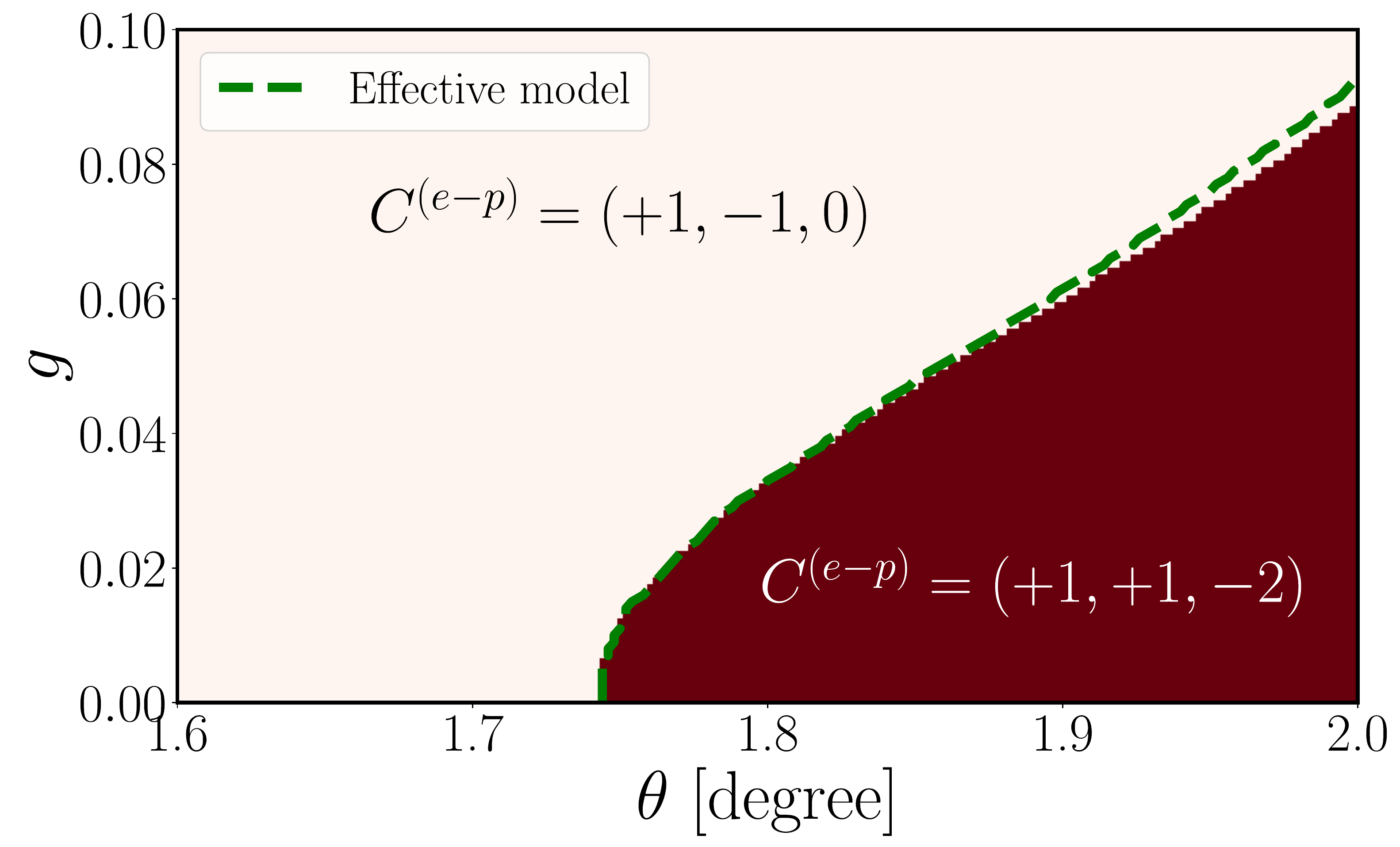}
\includegraphics[width = 0.95\hsize]{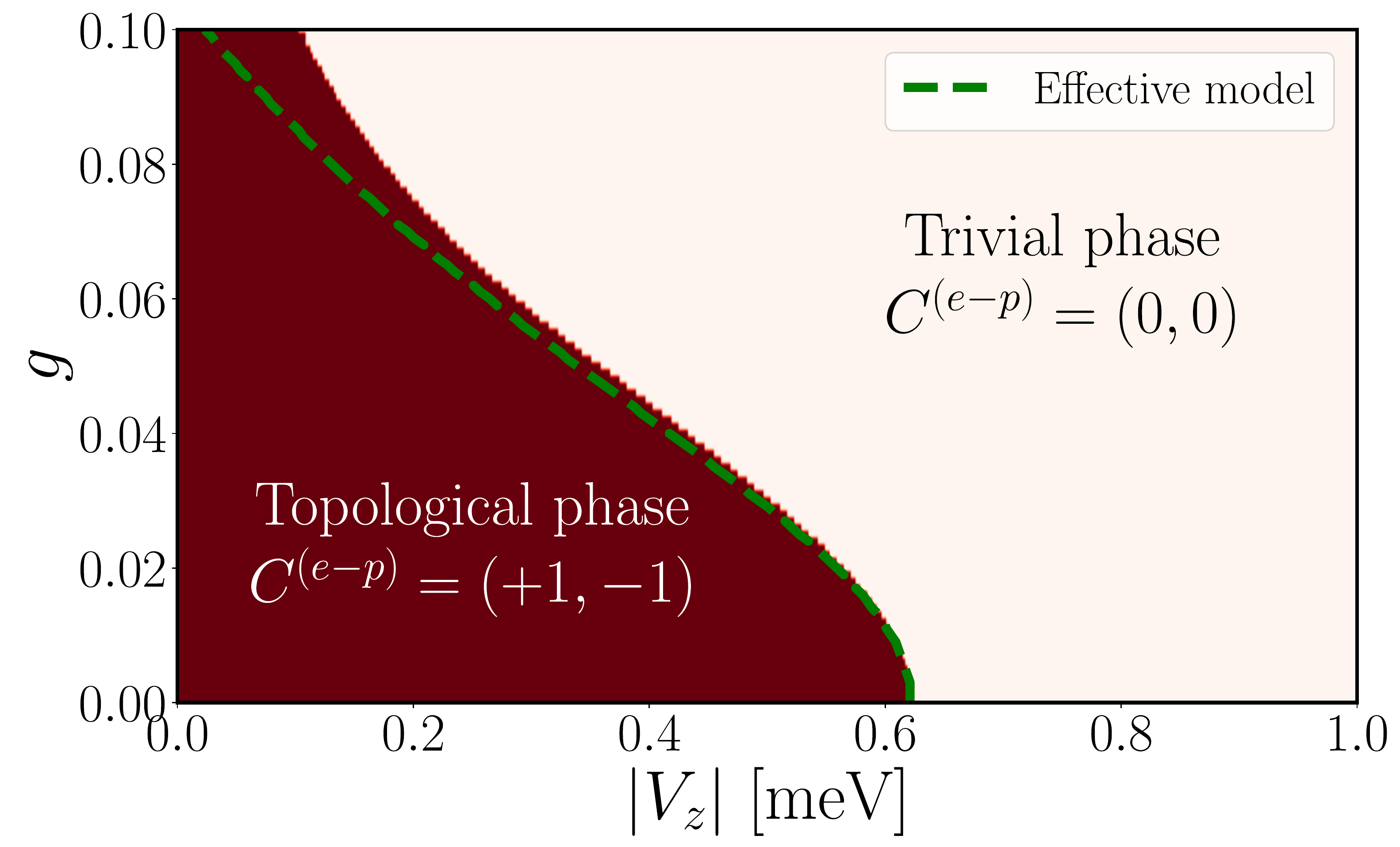}
  \caption{Electron-photon Chern numbers of the first three (top panel) and two (bottom panel) moir\'{e} minibands of the twisted TMD bilayer system (MoTe$_2$ parameters). The dashed lines depict the phase boundary  predicted by an effective electronic Hamiltonian obtained by adiabatically eliminating the photonic degrees of freedom \cite{Arwas2023} and calculating the electron Chern number. Top panel: Chern numbers versus the twisting angle $\theta$ and the dimensionless cavity coupling $g$ (see definition in the text), with no inter-layer bias ($V_z = 0$). Bottom panel: Chern numbers versus $V_z$ and the dimensionless cavity coupling $g$ for a fixed angle $\theta = 1.2^{\circ}$. Parameters: cavity photon energy $\hbar\omega_c = 20$ meV (top panel) and $6$ meV (bottom panel), cavity mode polarization vector $\mathbf{u} = (1,0,0)$.}
\label{fig:TMD-phase diagram}
\end{figure}

{\it Definition of the electron-photon Chern number ---} 
Let us consider a single fermion interacting with a cavity quantized field. This is equivalent to injecting a fermion into empty minibands.  If the fermion wavevector $\mathbf{k}$ is a good quantum number, then we can diagonalize the Hamiltonian  $\hat{\mathcal{H}}_{\mathbf{k}} = e^{-i\mathbf{k}\cdot\hat{\mathbf{r}}}\hat{\mathcal{H}}e^{i\mathbf{k}\cdot\hat{\mathbf{r}}}$ and obtain electron-photon eigenstates of the form $\vert \Psi^{{\mathrm{(e-p})}}_{n\mathbf{k}}\rangle$ with  corresponding electron-photon energy bands $\mathcal{E}^{{\mathrm{(e-p})}}_{n\mathbf{k}}$. Given the form of the system eigenstates, we can introduce the following electron-photon Chern number:
\begin{figure}[t!]
\includegraphics[scale = 0.23]{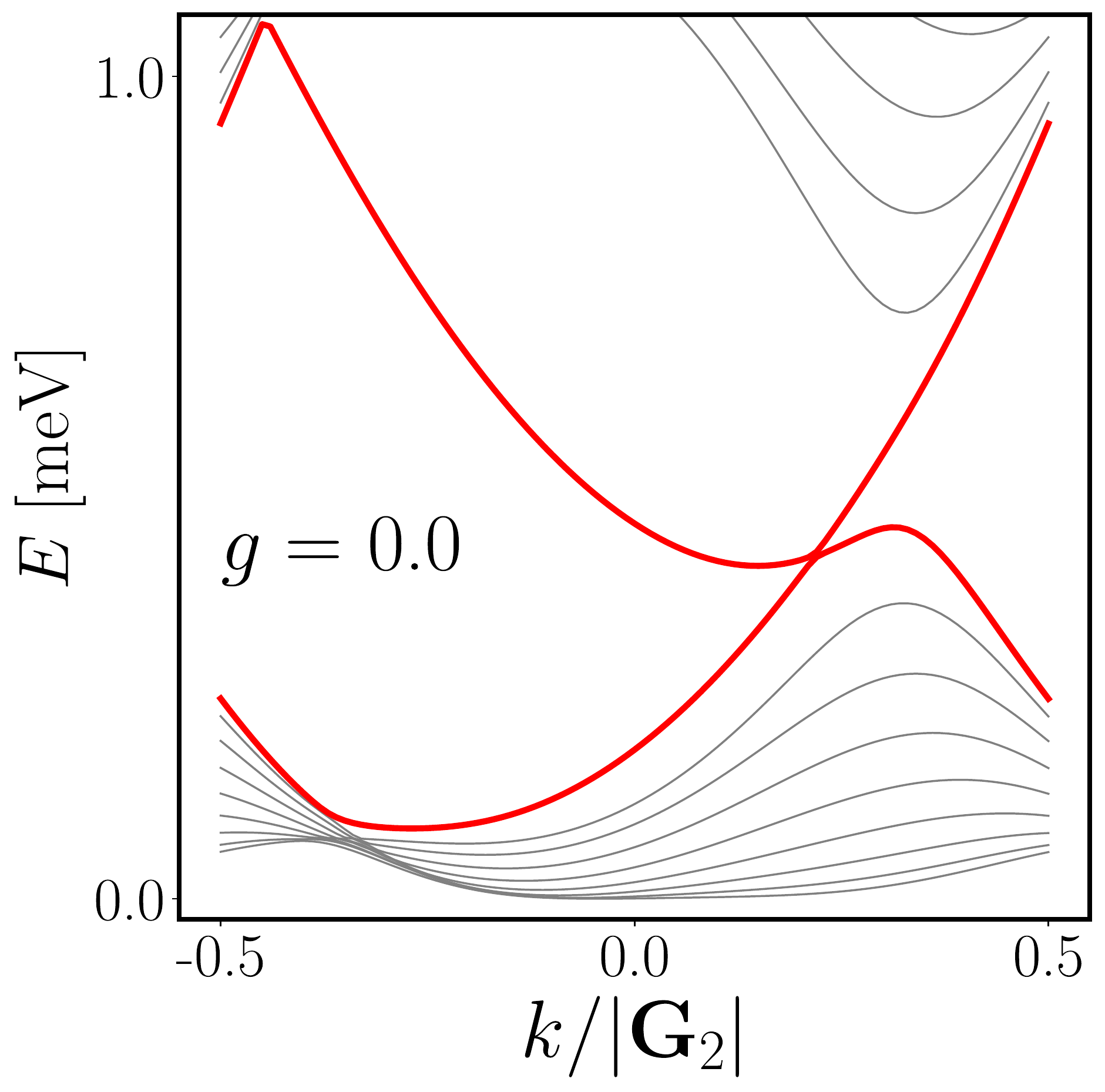}
\includegraphics[scale = 0.23]{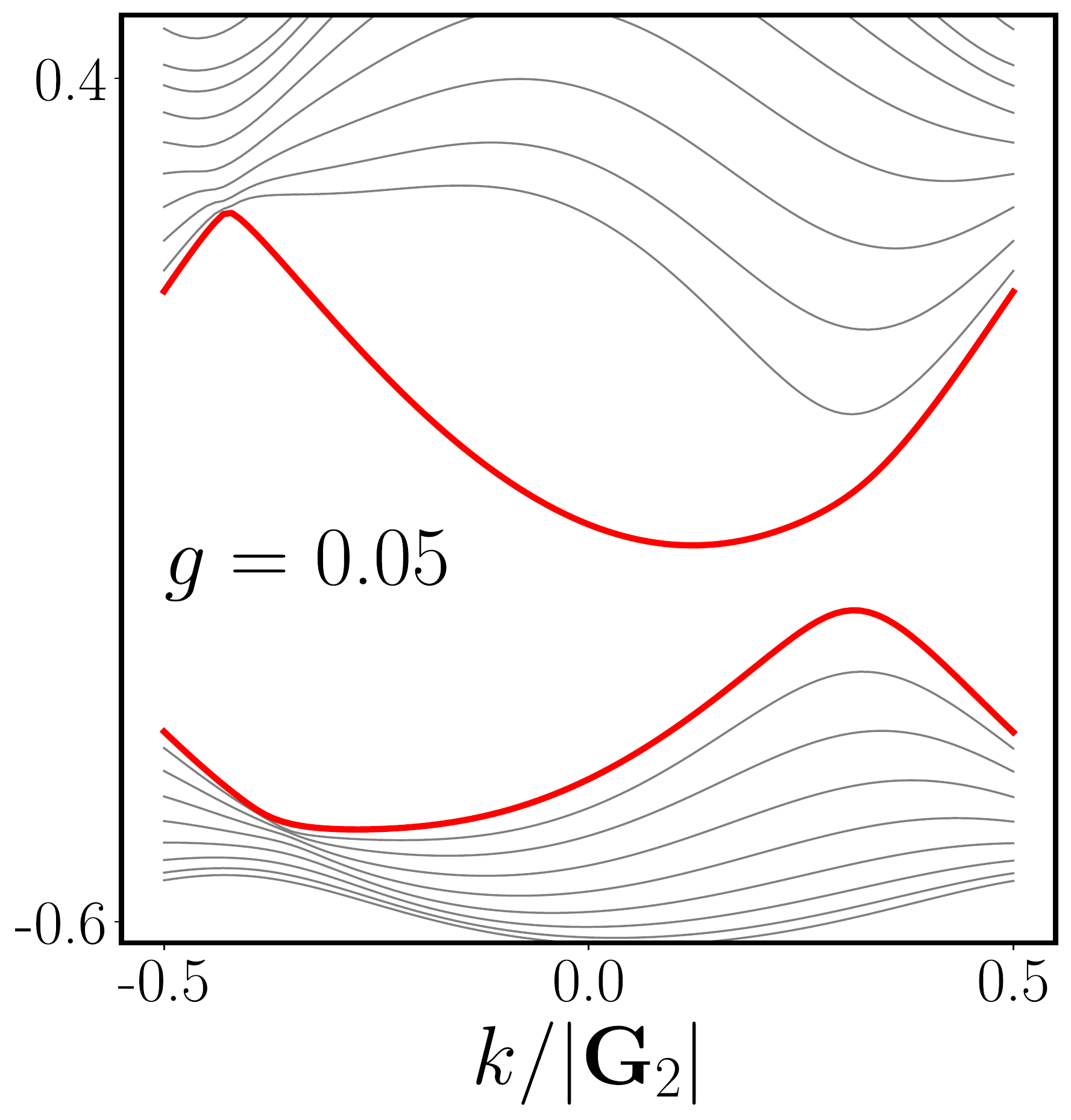}
  \caption{Electron-photon energy-momentum dispersions $\mathcal{E}^{(e-p)}_{n \mathbf{k}}$ for $g = 0.0$ (left panel) and $0.05$ (right panel) for a ribbon geometry (periodic boundary conditions along the long direction). Parameters: cavity photon energy $\hbar\omega_c = 6$ meV, $V_z = 0.5$ meV, $\theta = 1.2^{\circ}$, ribbon width $D_1/a_M = 10$ and cavity mode polarization vector $\mathbf{u} = (1,0,0)$. Here, the photonic component of the electron-photon eigenstates is small (see text).}
\label{fig:TMD-off res band}
\end{figure}
\begin{equation}
    \mathcal{C}^{{\mathrm{(e-p})}}_n = \int\frac{d^2k}{2\pi}i\sum_{\mu,\nu}\epsilon_{\mu\nu}\braket{\smash{\partial_{k_{\mu}}\Psi^{{\mathrm{(e-p})}}_{n\mathbf{k}}}}{\smash{\partial_{k_{\nu}}\Psi^{{\mathrm{(e-p})}}_{n\mathbf{k}}}},
    \label{eq:Cn}
    \end{equation}
where $\epsilon_{\mu\nu}$ is the two-dimensional Levi-Civita tensor. Importantly, the electron-photon Chern number in Eq. (\ref{eq:Cn}) is well defined for any arbitrary hybridization between the single electron and the cavity quantum field. If we work in the hole picture, the particle-hole transformation results in a simple change of sign of such Chern number. Numerically the Chern number in Eq. (\ref{eq:Cn}) has been calculated by numerical diagonalization of the Hamiltonian (\ref{eq:TMD-H1}) and using the technique reported in Ref. \cite{Zhao2020}.

{\it Results and discussions ---} In what follows, we
use the electron-photon Chern number to investigate topological properties of cavity-embedded moir\'e systems, focusing on twisted MoTe$_{2}$ bilayers. We first consider the scenario of high photon frequency, where the photon is off-resonant with respect to the relevant miniband electronic transitions. In Fig. \ref{fig:TMD-phase diagram}, we report the electron-photon topological Chern numbers associated to the first three moir\'e minibands of the twisted bilayer TMD system. Here, we consider a cavity mode with in-plane polarization $\mathbf{u} = (1,0,0)$. For $g = 0$ (no cavity coupling), the system is known to exhibit a topological transition as a function of the twisting angle $\theta$ (top panel) and as a function of the inter-layer bias $V_z$ (bottom panel). In the top panel we consider the case with no bias ($V_z = 0$) for which a topological transition occurs by increasing the twisting angle $\theta$ above a critical value ($\approx 1.75^{\circ}$ at zero $g$), with Chern numbers changing from $(+1,-1,0)$ to $(+1,+1,-2)$. The bottom panel corresponds to a situation with a fixed angle $\theta = 1.2^{\circ}$ for which a transition from topologically nontrivial Chern numbers $(+1,-1,0)$ to the trivial $(0,0,0)$  is achieved when $V_z$ is increased above a critical value ($\approx 0.63$ meV at zero $g$). 
\begin{figure}[t!]
\includegraphics[width = 0.9\hsize]{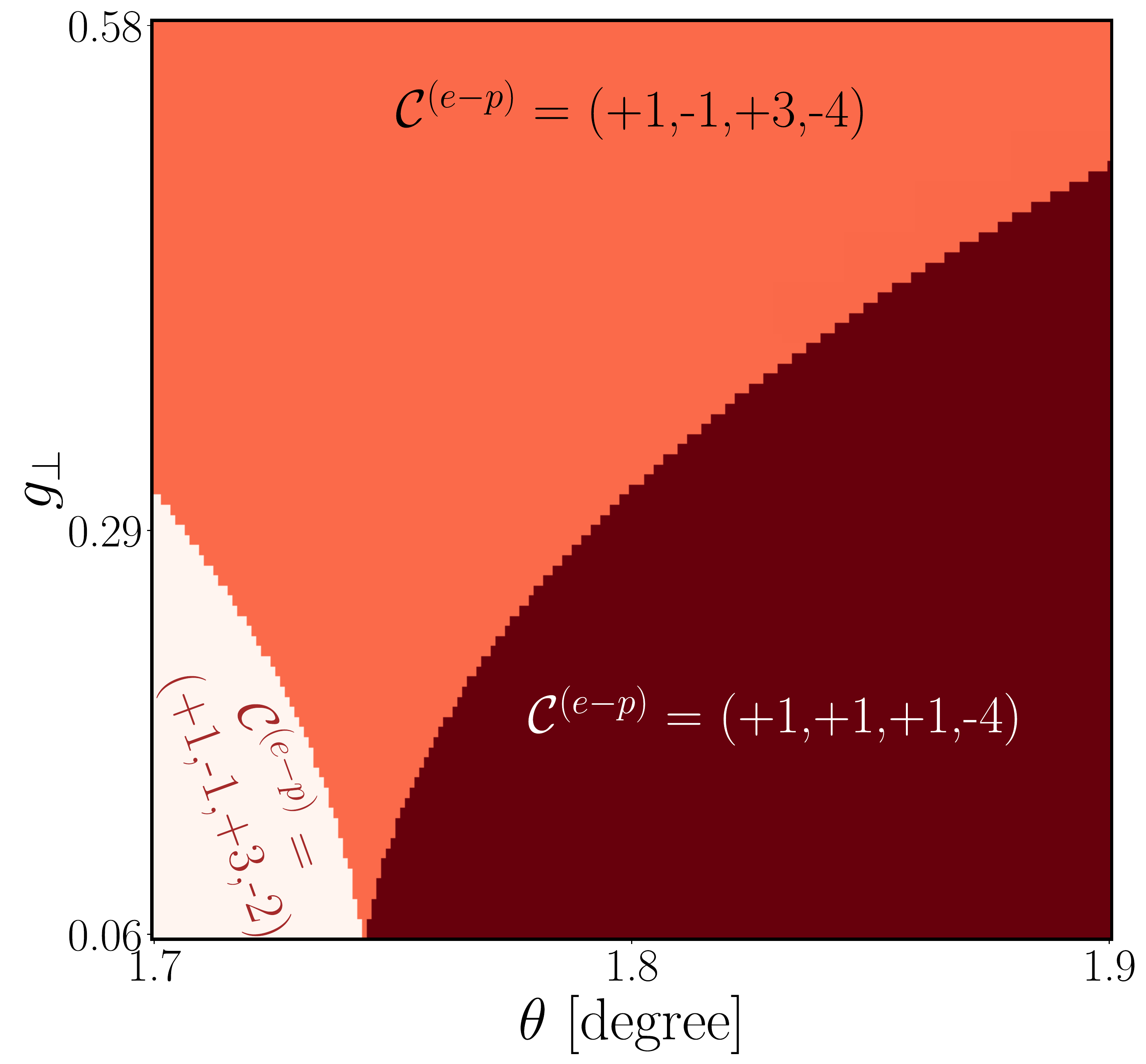}
  \caption{Electron-photon Chern numbers of the first four moir\'{e} minibands for the considered cavity-embedded twisted TMD bilayer system. Note that this topological diagram cannot be predicted by the effective electronic model used in Fig. \ref{fig:TMD-phase diagram}. Other parameters: cavity photon energy $\hbar\omega_c = 10$ meV, $V_z = 0$ meV, and cavity mode polarization $\mathbf{u} = (0,0,1)$.}
\label{fig:TMD-on phase diagram}
\end{figure}
In both situations considered in Fig. \ref{fig:TMD-phase diagram}, a finite cavity coupling modifies the transition boundary significantly for a range of dimensionless coupling $g$, which is accessible with deeply sub-wavelength resonators. The energy-momentum dispersions with and without the cavity are compared for $V_z = 0.5$ meV in Fig. \ref{fig:TMD-off res band} for a ribbon geometry (periodic boundary conditions along the long direction). The left panel shows the presence of topological edge states that cross in energy for $g=0$, while the right panel shows the opening of an edge gap in the presence of a cavity with $g = 0.05$. In other words, Fig. \ref{fig:TMD-off res band} shows the consequence of the topological transition with increasing cavity coupling $g$ depicted in the diagram of the bottom panel in Fig. \ref{fig:TMD-phase diagram}. 

Note that by tracing out the photonic degrees of freedom, we can obtain an electronic reduced density matrix, namely $\hat{\rho}_{n\mathbf{k}} = \text{Tr}_{\text{phot}}\left(\ket{\smash{\Psi_{n\mathbf{k}}^{(e-p)}}}\bra{\smash{\Psi_{n\mathbf{k}}^{(e-p)}}}\right)$. The electronic purity of such reduced density matrix is defined as ${\mathbb P}_{n\mathbf{k}} = \text{Tr}_{\text{el}}(\hat{\rho}_{n\mathbf{k}}^2)$ \cite{CohenTannoudji2020} and is equal to $1$ for a purely electronic state. For $g = 0.1$, the minimum of the electronic purity (the purity depends on $\mathbf{k}$) is about $75\%$ for the top panel of Fig. \ref{fig:TMD-off res band} and $88\%$ for the bottom one. In such configuration, one might describe the system with an effective electronic Hamiltonian \cite{Arwas2023,Dmytruk2022,Wang2019,Li2022,Guerci_PRL_2020}. Indeed, we as shown in of Fig. \ref{fig:TMD-phase diagram} the diagrams are qualitatively reproduced by an effective electronic Hamiltonian approach (see Supplementary Material).
However our electron-photon Chern number introduced in Eq. (\ref{eq:Cn}) and based on the exact light-matter energies $\mathcal{E}^{(e-p)}_{n \mathbf{k}}$ is defined also for low electronic purity, i.e., for any arbitrary light-matter hybridization. 

\begin{figure}[t!]
\includegraphics[scale=0.235]{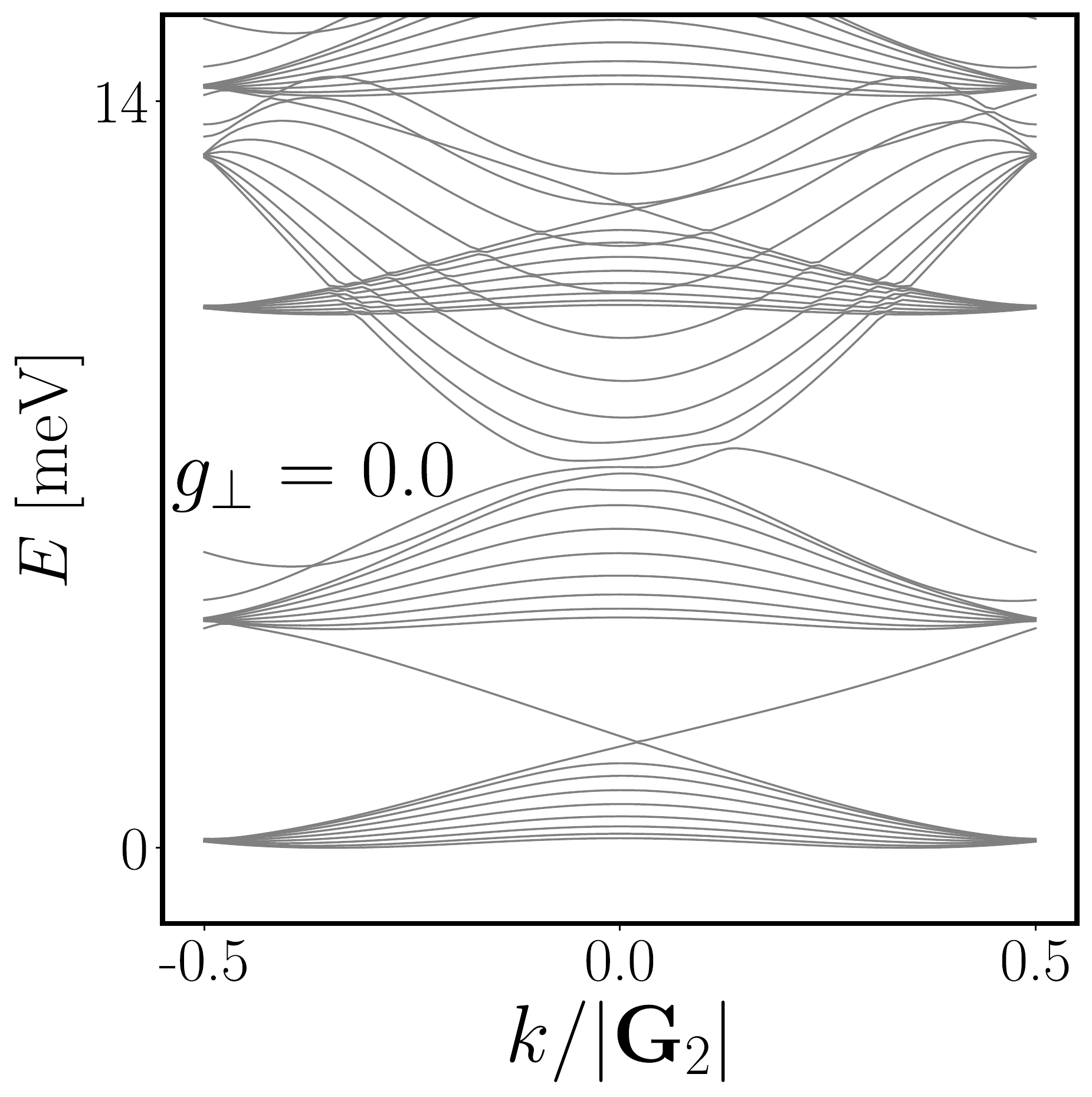}
\includegraphics[scale=0.235]{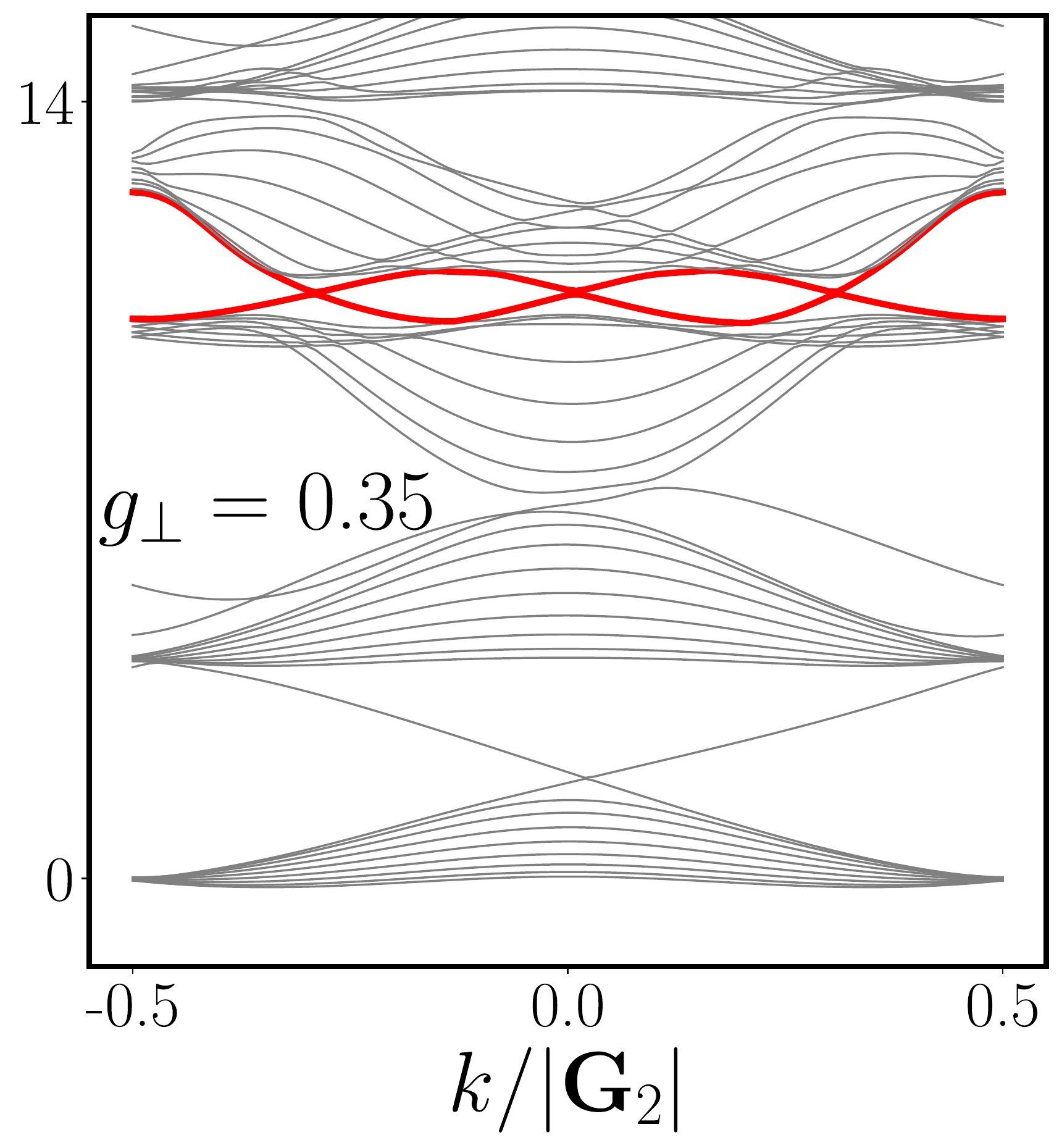}\\
\includegraphics[scale=0.235]{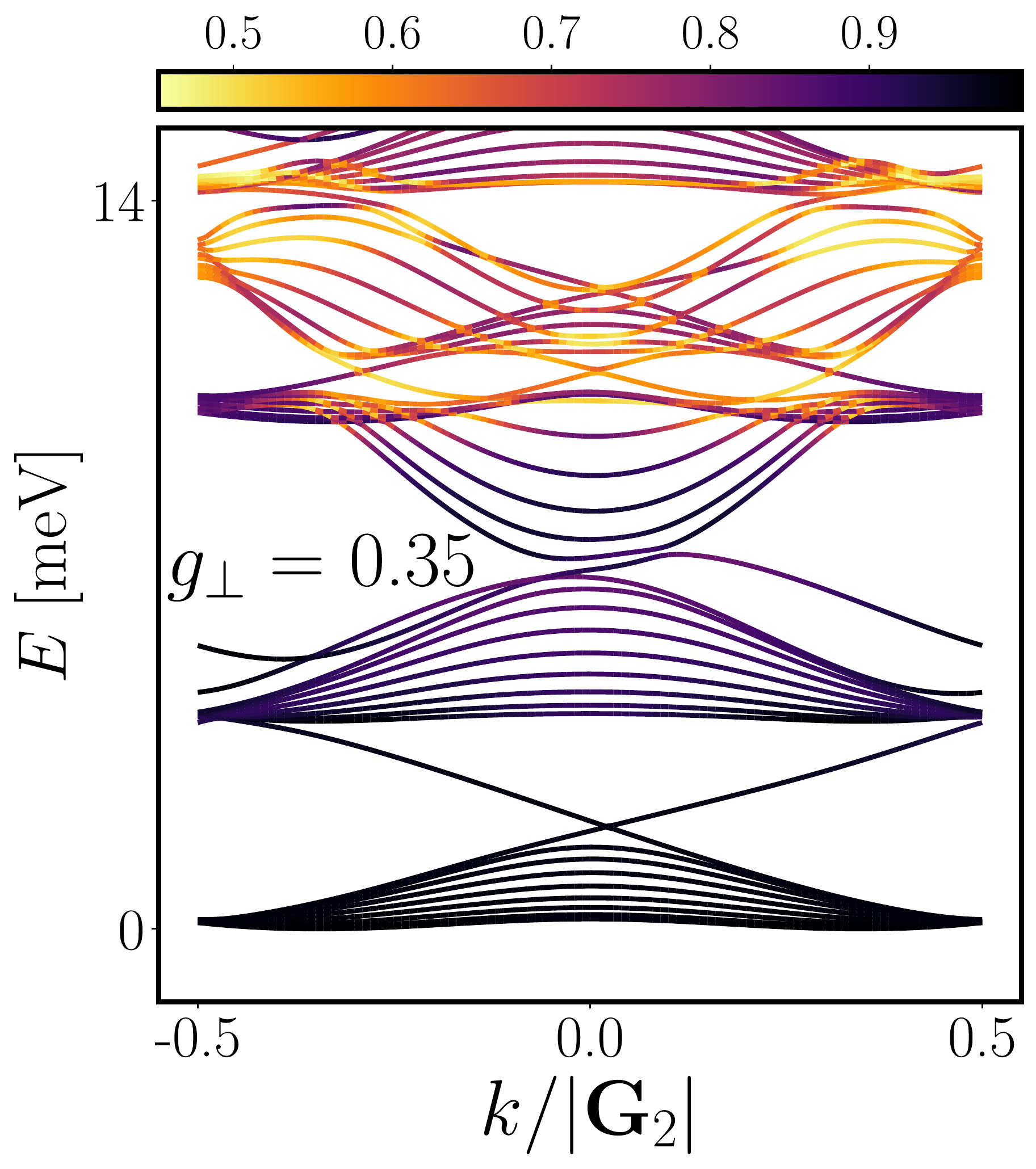}
\includegraphics[scale=0.235]{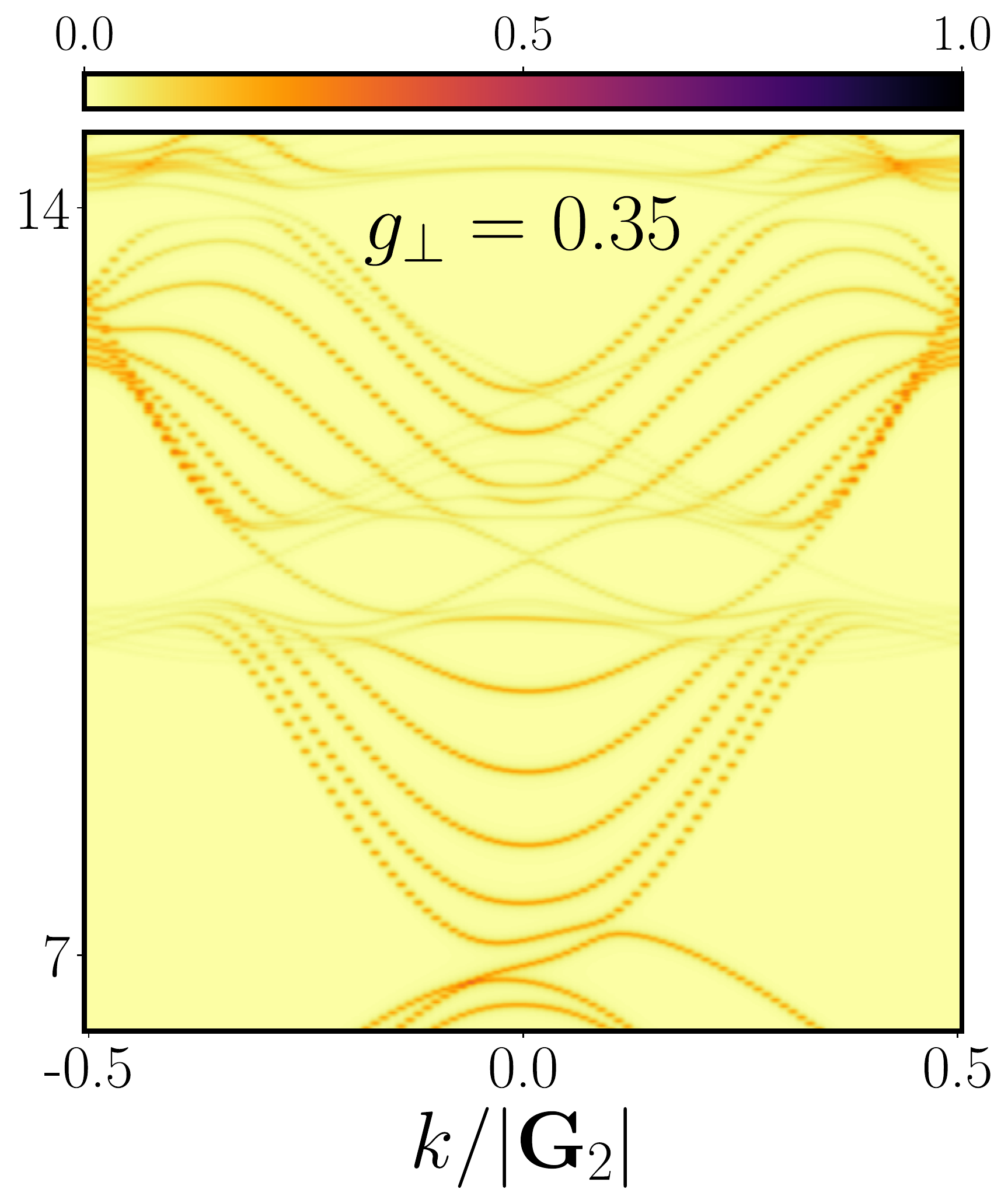}
  \caption{Top panels: electron-photon miniband dispersions $\mathcal{E}^{(e-p)}_{n \mathbf{k}}$ for $g_{\perp} = 0$ (left panel) and $0.35$ (right panel), where three edge states emerge (thicker red lines). Bottom panels: electronic purity (left panel) and electronic spectral function (right panel, normalized by its maximum value)  for $g_{\perp} = 0.35$. We have considered a ribbon geometry with width $D_1/a_M = 10$ and periodic boundary conditions along the long direction. Other parameters: cavity photon energy $\hbar\omega_c = 10$ meV, $V_z = 0$ meV, $\theta = 1.8^{\circ}$, cavity mode polarization $\mathbf{u} = (0,0,1)$. }
\label{fig:res-band}
\end{figure}
\label{discuss}

\label{res}
Let us now consider the situation where the cavity mode has a relatively low frequency and is resonant with electronic miniband transitions. Note that we consider here out-of-plane cavity polarization $\mathbf{u} = (0,0,1)$ and the corresponding dimensionless coupling constant is defined as $g_{\perp} = eA_0d/\hbar$. The corresponding results are depicted in Fig. \ref{fig:TMD-on phase diagram} and \ref{fig:res-band}. In the non-interacting case ($g_{\perp}=0$), the photon energy creates replicas of the electronic minibands with non-zero photon numbers, as shown in Fig. \ref{fig:res-band}. The bare electron-photon spectrum becomes degenerate when a photon replica crosses the original electronic miniband with zero photons.
However, a finite coupling lifts such degeneracies, leading to new electron-photon energy minibands for which we can compute the electron-photon Chern number.
Fig. \ref{fig:TMD-on phase diagram} shows the topological diagram characterized by such an electron-photon Chern numbers for the first four minibands. The cavity photon energy is resonant to the transition  between the first and third electronic minibands, depicted in Fig. \ref{fig:res-band}.  The bulk-edge correspondence for the electron-photon Chern number is  satisfied: indeed, we observe three edge states shown by the thicker red line in Fig. \ref{fig:res-band}. These states have low electronic purity (as low as 50\%) and cannot be captured by an effective electronic Hamiltonian where the photon degrees of freedom are eliminated. Moreover, such high Chern numbers, created by the hybridization with the cavity photon replicas, are absent in the bare electronic system. These multiple new edge states corresponding to high Chern numbers can be observed by measuring the electronic spectral function via Angle-Resolved Photoemission Spectroscopy (ARPES) \cite{Noguchi2021}, Scanning Tunneling Microscopy (STM) \cite{Tao2011} or Microwave Impedance Microscopy \cite{Barber2021}.

{\it Conclusions ---} In this letter, we explored the new topological phases characterized  by the electron-photon Chern number, a topological invariant defined in terms of the exact eigenstates for the Hamiltonian describing a fermionic particle coupled to a quantized electromagnetic field. This is a topological invariant for any arbitrary hybridization of the electron-photon eigenstates. 
Note that electron-photon Chern numbers can also defined for disordered and non-homogeneous systems \cite{Prodan2010,Bianco2011,Zhang2013,Caio2019}, so the present topological approach can also be generalized to situations where the cavity mode is not spatially homogeneous or in the presence of electronic disorder. The results of this letter are exact when a single fermion is injected in empty minibands. A future and intriguing problem to explore is how these topological states evolve when such electron-photon states are partially filled.

\acknowledgements
{We acknowledges financial support from the French agency ANR through the project NOMOS (ANR-18-CE24-0026), and TRIANGLE (ANR-20-CE47-0011), CaVdW under the ANR/RGC Joint Research Scheme sponsored by ANR and RGC of Hong Kong SAR China (ANR-21-CE30-0056-01, A-HKU705/21), and from the Israeli Council for Higher Education - VATAT. 
Z.L. and W.Y. also acknowledge support by RGC of Hong Kong SAR (HKU SRFS2122-7S05, AoE/P-701/20).
}

\bibliography{Bib.bib}

\end{document}


\title{{\bf Supplementary Material for the article:}\\ ``Electron-photon Chern number in cavity-embedded 2D moir\'{e} materials"}

\author{Danh-Phuong Nguyen}
\affiliation{Universit\'{e} Paris Cit\'e, CNRS, Mat\'{e}riaux et Ph\'{e}nom\`{e}nes Quantiques, 75013 Paris, France}
\author{Geva Arwas}
\affiliation{Universit\'{e} Paris Cit\'e, CNRS, Mat\'{e}riaux et Ph\'{e}nom\`{e}nes Quantiques, 75013 Paris, France}
\author{Zuzhang Lin}
\affiliation{Department of Physics, The University of Hong Kong, Hong Kong, China}
\affiliation{HKU-UCAS Joint Institute of Theoretical and Computational Physics at Hong Kong, China} 
\author{Wang Yao}
\affiliation{Department of Physics, The University of Hong Kong, Hong Kong, China}
\affiliation{HKU-UCAS Joint Institute of Theoretical and Computational Physics at Hong Kong, China}
\author{ Cristiano Ciuti}
\affiliation{Universit\'{e} Paris Cit\'e, CNRS, Mat\'{e}riaux et Ph\'{e}nom\`{e}nes Quantiques, 75013 Paris, France}

\date{\today}
\maketitle
\newtagform{supplementary}[]()
\section{Cavity QED Hamiltonian}
In this section we derive a four-band continuum model \cite{Wu2019,Zhai2020} for small angle twisted bilayer TMD. The electron-photon Hamiltonian can be written as:
\begin{equation}
\label{eq:S-f4}
    \hat{H} = \hbar\omega_c\hat{a}^{\dagger}\hat{a} + \begin{pmatrix}
        \hat{H}_{t} & 0 \\ 0 & \hat{H}_{b}
    \end{pmatrix} + \begin{pmatrix}
        \hat{V}_t & \hat{U} \\ \hat{U}^{\dagger} & \hat{V}_b
    \end{pmatrix},
\end{equation}
where $\hat{a}^{\dagger}$ creates a photon of frequency $\omega_c$, $\hat{H}_{t}$ ($\hat{H}_{b}$) is a $2 \cross 2$ matrix corresponding to the top (bottom) layer describing the TMD material along with the coupling to the cavity photon. The last term is a $4 \cross 4$ matrix describing the moir\'{e} potential. $\hat{V}_{t}$ and $\hat{V}_{b}$ describe the intra-layer interactions, while $\hat{U}$ is the layer-layer interaction. Each layer is a honeycomb lattice with the lattice constant $a_0$ and is defined by the two primitive vectors $\mathbf{a}_1 = a_0\sqrt{3}/2(1, \sqrt{3}, 0)$, $\mathbf{a}_2 = a_0\sqrt{3}/2(-1, \sqrt{3}, 0)$ and their reciprocal counterparts $\mathbf{b}_1 = 2\pi/(3a_0)(\sqrt{3},1,0)$, $\mathbf{b}_2 = 2\pi/(3a_0)(-\sqrt{3},1,0)$. 
The top layer is rotated in-plane by a small angle $\theta$ in order to create a moir\'{e} pattern. 
The mismatch between layers is given by $\boldsymbol{\Delta}(\mathbf{r}) = [\mathbb{1} - R^{-1}(\theta)]\mathbf{r}$, where $R(\theta)$ is a rotation matrix about the $z$-axis. One moir\'{e} unit cell is defined by $\mathbf{L}_{i=1,2}$ such that $\boldsymbol{\Delta}(\mathbf{L}_i)= \mathbf{a}_i$, hence $\mathbf{L}_i = [\mathbb{1} - R^{-1}(\theta)]^{-1}\mathbf{a}_i$. Similarly, the moir\'{e} reciprocal lattice vectors $\mathbf{G}_i = [\mathbb{1} - R(\theta)]\mathbf{b}_i$, satisfying $\mathbf{L}_i \cdot \mathbf{G}_j = 2\pi \delta_{ij}$. 
\begin{figure}
\centering
\includegraphics[width = 0.49\hsize]{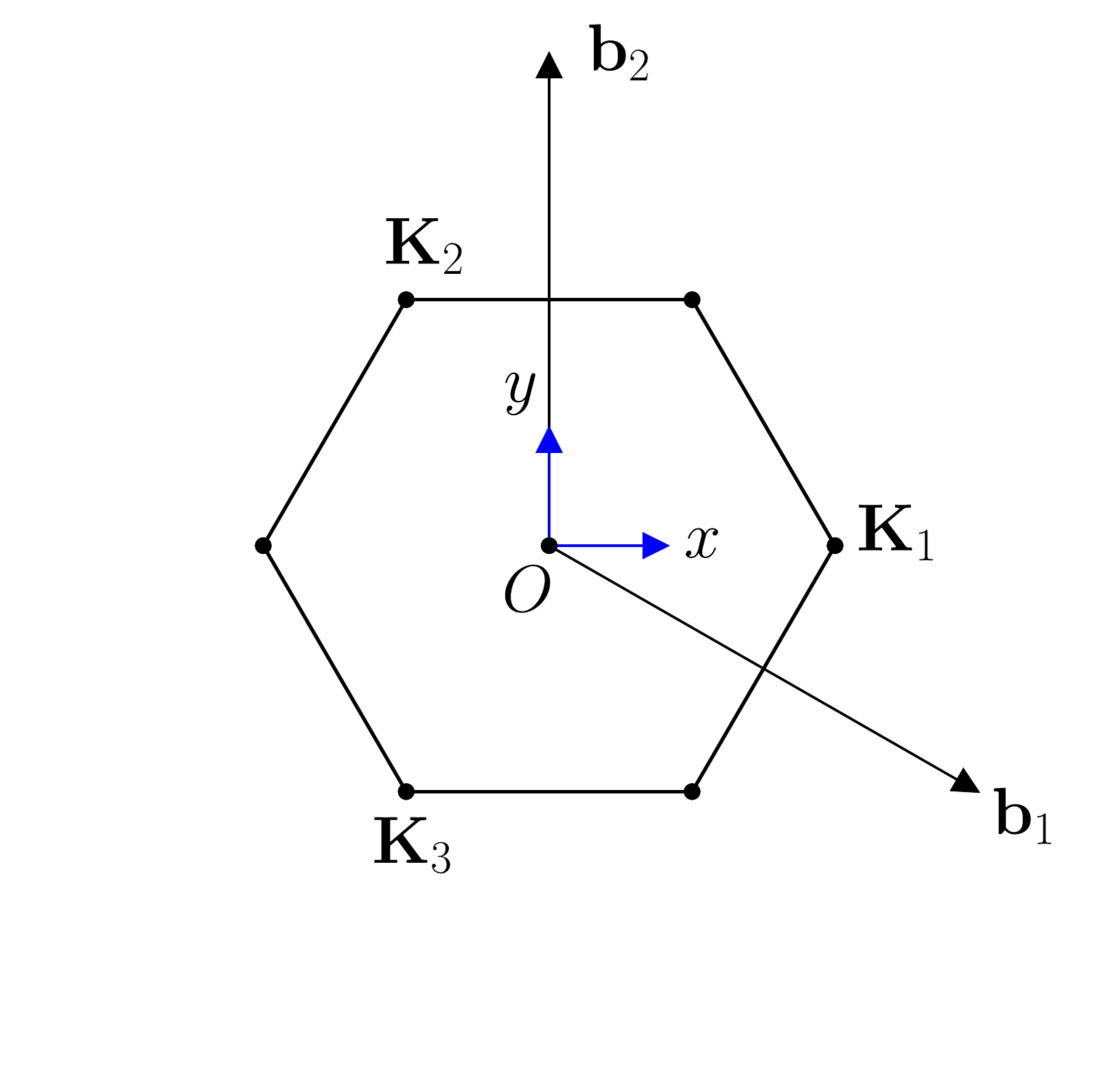}
\includegraphics[width = 0.49\hsize]{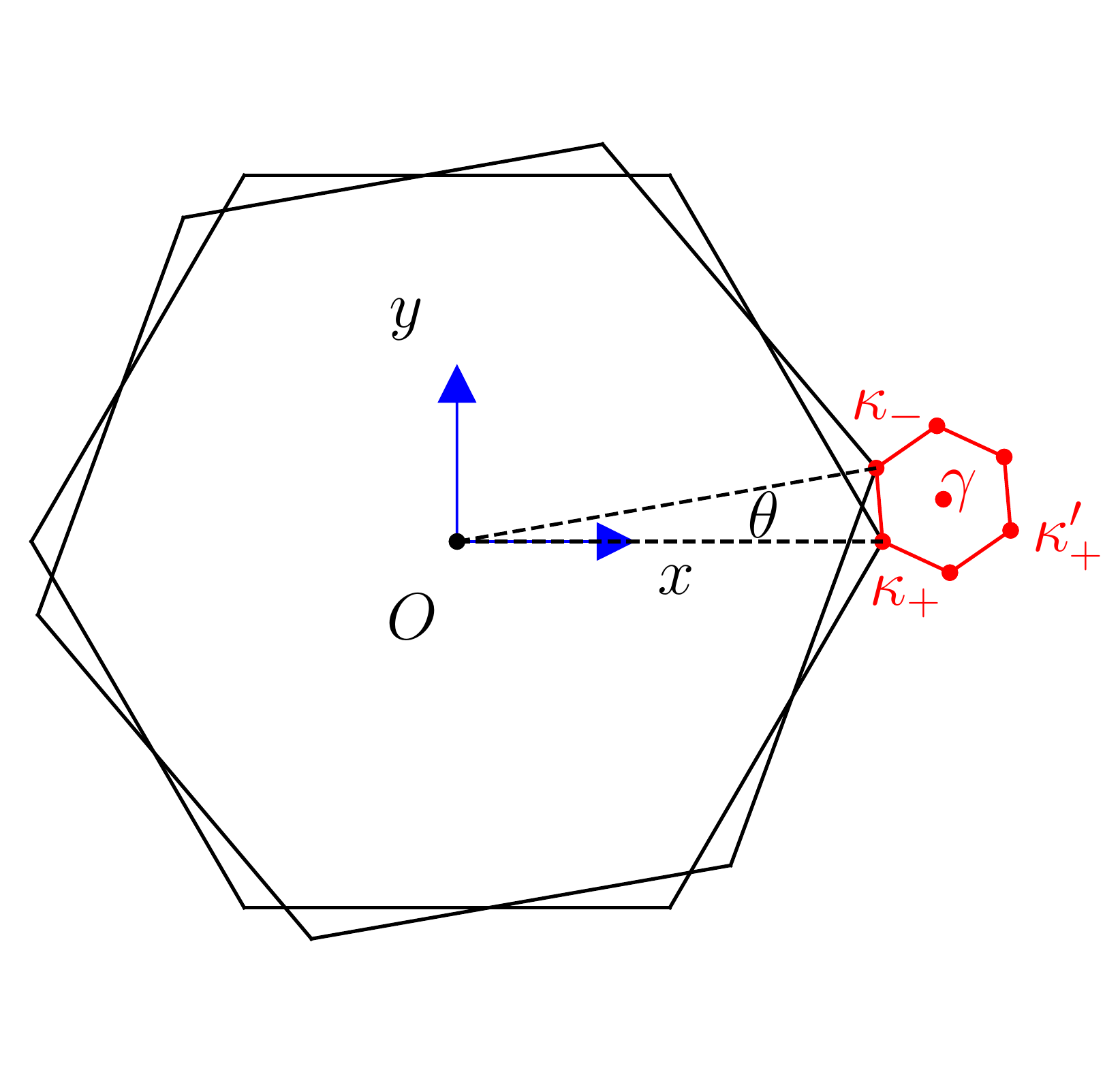}
\caption{First Brillouin zone with Dirac points of the bottom layer (left) and moir\'{e} Brillouin zone (shown in red) of moir\'e superlattice (right).}\label{fig:SM-TMD-mBZ}
\end{figure}

Each layer is described by a tight binding Hamiltonian on the honeycomb lattice, with hopping coupling quantified by $t$. The unit cell consists of two sites $A, B$ having an on-site energy difference of $2\Delta$. The electron-photon interaction is introduced via Peierls substitution. For the bottom layer, we can write the Hamiltonian as:
\begin{equation}
\label{eq:S:H2f}
    \begin{aligned}
        \hat{H}_{b} &= \Delta\sum_{\mathbf{R}}\bigg(\ket{A,\mathbf{R}}\bra{A,\mathbf{R}} - \ket{B,\mathbf{R}}\bra{B,\mathbf{R}}\bigg)\\
        &+ t\sum_{\mathbf{R},j=1,2,3}e^{-ig(\hat{a}+\hat{a}^{\dagger})\mathbf{u}\cdot\mathbf{e}_j}\ket{B,\mathbf{R} + \mathbf{a}_j}\bra{A,\mathbf{R}} + \text{h.c},
    \end{aligned}
\end{equation}
where $g = eA_0a_0/\hbar$, $\mathbf{u}$ is the cavity orientation, $\mathbf{e}_{j=1,2,3} = R\left(j2\pi/3\right)(0,-1,0)$ and $\mathbf{a}_3 = (0,0,0)$. We assume the cavity field $\mathbf{A}$ is uniform, such that the Hamiltonian (\ref{eq:S:H2f}) is block-diagonal in $\mathbf{k}$ space. The Bloch Hamiltonian $H^e_b({\mathbf{k}})$ is given by:
\begin{equation}
\label{eq:S:H3f}
    \begin{aligned}
        \hat{H}_b(\mathbf{k}) &= \Delta\ket{A,\mathbf{k}}\bra{A,\mathbf{k}} - \Delta\ket{B,\mathbf{k}}\bra{B,\mathbf{k}}\\
        &+ t\sum_{j=0,1,2} e^{-ig(\hat{a}+\hat{a}^{\dagger})\mathbf{u}\cdot\mathbf{e}_j}e^{-i\mathbf{k}\cdot\mathbf{a}_j}\ket{B,\mathbf{k}}\bra{A,\mathbf{k}} + \text{h.c},
    \end{aligned}
\end{equation}
where $\mathbf{k}$ is in the Brillouin zone (BZ) as shown in Figure \ref{fig:SM-TMD-mBZ}. Next, we expand around one valley $\mathbf{K}_b$ such that $\mathbf{K}_b\cdot\mathbf{a}_1 = -2\pi/3$, and at the  first order in $g$, we have: 
\begin{equation}
\begin{aligned}
\hat{H}_b({\mathbf{k}}) &= \Delta\sigma_z + \frac{3a_0t}{2}\left(\mathbf{k} - \mathbf{K}_b + \frac{e\mathbf{A}^{(xy)}}{\hbar}\left(\hat{a} + \hat{a}^{\dagger}\right)\right)\cdot \boldsymbol{\sigma},
\end{aligned}
\label{SM:eq-Heb}
\end{equation}
where $\boldsymbol{\sigma} = (\sigma_x,\sigma_y,\sigma_z)$ are the Pauli matrices for the sub-lattice degree of freedom, $\mathbf{A}^{(xy)}$ is the in-plane projection of the cavity field $\mathbf{A}$.

The top layer is rotated in real space by an angle $\theta$ such that the lattice vectors in equation (\ref{eq:S:H2f}) are transformed as $\mathbf{a} \to \tilde{\mathbf{a}} = R(\theta)\mathbf{a}$. Following the same procedure and restricting $\mathbf{k}$ to the vicinity of $\mathbf{K}_t = R(\theta)\mathbf{K}_b$ one gets:
\begin{equation}
\label{SM:eq-Het}
\begin{aligned}
\hat{H}_t({\mathbf{k}}) \ &= \ \Delta\sigma_z \\ &+ \frac{3a_0t}{2}R^{-1}(\theta)\left(\mathbf{k} - \mathbf{K}_t + \frac{e\mathbf{A}^{(xy)}}{\hbar}\left(\hat{a} + \hat{a}^{\dagger}\right)\right)\cdot \boldsymbol{\sigma}.
\end{aligned}
\end{equation}

In the presence of the cavity, the interlayer hopping $\hat{U}$ between site $\mathbf{R}_b$ of the bottom and $\mathbf{R}_t$ of the top can be approximated as:
\begin{equation}
    \hat{U} = \sum_{\substack{\mathbf{R}_b,\mathbf{R}_t \\ \sigma = A,B}}e^{-i\Phi_{\mathbf{R}_t,\mathbf{R}_b}(\hat{a}+\hat{a}^{\dagger})}T(\mathbf{R}_t -\mathbf{R}_b)\ket{\sigma,\smash{\mathbf{R}_t}}\bra{\sigma,\mathbf{R}_b} ,
\end{equation}
where $ T(\mathbf{R}_t -\mathbf{R}_b)$ is the hopping coefficient in the absence of a cavity \cite{Wang2017,Bistritzer2011}. We assume that the interlayer hopping is only between sub-lattices of the same type. $\mathbf{R}_t - \mathbf{R}_b = \mathbf{d} + \boldsymbol{\Delta}(\mathbf{r})$ where $\mathbf{d} = d(0,0,1)$ is the distance between two layers. The Peierls phase can be approximated by $\Phi_{\mathbf{R}_t,\mathbf{R}_b} = e\mathbf{A}\cdot(\mathbf{R}_t-\mathbf{R}_b)/\hbar \approx eA^{(z)}d/\hbar$, where $A^{(z)}$ is the $z$-component of the cavity field. Since the Peierls phase does not depend on $\mathbf{r}$, we can write the interlayer coupling as:
\begin{equation}
    \hat{U} = e^{-i\frac{eA^{(z)}d}{\hbar}(\hat{a}+\hat{a}^{\dagger})}\hat{U}_0,
\end{equation}
where $\hat{U}_0$ is a moir\'{e} interlayer hopping in the absence of the cavity. 

Combining all terms and defining $v_F = 3a_0 t/(2\hbar)$ one gets a low-energy Hamiltonian for the twisted bilayer TMD coupled to cavity field:
\begin{widetext}
\begin{equation}
\label{eq:S:H5}
    \widetilde{H} = 
    \begin{pmatrix}
    \Delta \sigma_z + v_F R^{-1}(\theta)\left[\mathbf{\hat{p}} - \hbar\boldsymbol{\kappa}_- +  e\mathbf{A}^{(xy)}(\hat{a}+\hat{a}^{\dagger}) \right]\cdot\boldsymbol{\sigma} + \hat{V}_t & e^{-i\frac{eA^{(z)} d}{\hbar} (\hat{a}+\hat{a}^{\dagger})}\hat{U}_0 \\
    e^{i\frac{eA^{(z)} d}{\hbar}(\hat{a}+\hat{a}^{\dagger})}\hat{U}_0^{\dagger} &  \Delta \sigma_z + v_F \left[\mathbf{\hat{p}} -\hbar\boldsymbol{\kappa}_+ + e\mathbf{A}^{(xy)}(\hat{a}+\hat{a}^{\dagger})  \right]\cdot\boldsymbol{\sigma}  + \hat{V}_b
    \end{pmatrix} + \hbar \omega_c \hat{a}^{\dagger}\hat{a}.
\end{equation}
Note that here we change the notation as $\mathbf{K}_b$ ($\mathbf{K}_t$) $\rightarrow \boldsymbol{\kappa}_+ $ ($\boldsymbol{\kappa}_-$) to highlight the moir\'{e} mini-BZ (see Fig. \ref{fig:SM-TMD-mBZ}). Next, we apply the unitary transformation $\hat{T}_{\theta} = \text{diag}(e^{-i\frac{\theta}{2}\sigma_z},\mathbb{1})$, leading to
\begin{equation}
    \hat{T}_{\theta}\widetilde{H}\hat{T}_{\theta}^{\dagger} = \begin{pmatrix}
    \Delta \sigma_z + v_F \left[\mathbf{\hat{p}} - \hbar\boldsymbol{\kappa}_- +  e\mathbf{A}^{(xy)}(\hat{a}+\hat{a}^{\dagger}) \right]\cdot\boldsymbol{\sigma} + \hat{V}_t & e^{-i\frac{eA^{(z)} d}{\hbar} (\hat{a}+\hat{a}^{\dagger})}e^{-i\frac{\theta}{2}\sigma_z}\hat{U}_0 \\
    e^{i\frac{eA^{(z)} d}{\hbar}(\hat{a}+\hat{a}^{\dagger})}e^{i\frac{\theta}{2}\sigma_z}\hat{U}_0^{\dagger} &  \Delta \sigma_z + v_F \left[\mathbf{\hat{p}} -\hbar\boldsymbol{\kappa}_+ + e\mathbf{A}^{(xy)}(\hat{a}+\hat{a}^{\dagger})  \right]\cdot \boldsymbol{\sigma} + \hat{V}_b
    \end{pmatrix} + \hbar \omega_c \hat{a}^{\dagger}\hat{a} .
\end{equation}
We remove the photonic operators from the off-diagonal terms using the unitary transformation $\hat{T}_{d} = \text{diag}(e^{i\frac{eA^{(z)}d}{2\hbar}\left(\hat{a}+ \hat{a}^{\dagger}\right)},e^{-i\frac{eA^{(z)}d}{2\hbar}\left(\hat{a}+ \hat{a}^{\dagger}\right)})$. The photonic number operator transforms as $\hat{a}^{\dagger}\hat{a} \to \hat{a}^{\dagger}\hat{a} + i\frac{eA^{(z)}d}{2\hbar}\left(\hat{a} - \hat{a}^{\dagger}\right)\tau_z + \text{constant}$, where $\tau_z$ is the Pauli matrix with respect to the layer pseudo-spin. Finally, the four-band Hamiltonian $\hat{H}^{(4)} = \hat{T}_d\hat{T}_{\theta}\widetilde{H}\hat{T}^{\dagger}_{\theta}\hat{T}^{\dagger}_d$ can be written as:
\begin{equation}
\label{eq:S:Hf}
\begin{aligned}
    \hat{H}^{(4)} &= \begin{pmatrix}
    \Delta \sigma_z + v_F \left[\mathbf{\hat{p}} - \hbar\boldsymbol{\kappa}_- +  e\mathbf{A}^{(xy)}(\hat{a}+\hat{a}^{\dagger}) \right]\cdot\boldsymbol{\sigma} + \hat{V}_t & e^{-i\frac{\theta}{2}\sigma_z}\hat{U}_0 \\
    e^{i\frac{\theta}{2}\sigma_z}\hat{U}_0^{\dagger} &  \Delta \sigma_z + v_F \left[\mathbf{\hat{p}} -\hbar\boldsymbol{\kappa}_+ + e\mathbf{A}^{(xy)}(\hat{a}+\hat{a}^{\dagger})  \right]\cdot \boldsymbol{\sigma} + \hat{V}_b
    \end{pmatrix} \\
    &+ \hbar \omega_c \hat{a}^{\dagger}\hat{a} + i\frac{\omega_c e A^{(z)} d}{2}\left(\hat{a} - \hat{a}^{\dagger}\right)\tau_z.
    \end{aligned}
\end{equation}
\end{widetext}
\section{Effective model for the valence bands}

For the case of large $\Delta$ we can construct an effective model for the valence bands. Let $\Psi = (\Psi_{t}^c, \Psi_{t}^v, \Psi_{b}^c, \Psi_{b}^v)^T$ be an eigenvector of $\hat{H}^{(4)}$ with energy $E$, where $c$ ($v$) corresponds to the conduction (valence) bands. Using (\ref{eq:S:Hf}), we can write
\begin{widetext}
 \begin{equation}
 \label{SM:eq-all 4}
 \begin{aligned}
     \left(\Delta + \hat{V}_t^c + \hbar\omega_c\hat{a}^{\dagger}\hat{a} + \hat{D}\right)\Psi_{t}^c + v_F\left(\hat{\Pi}_{tx} - i\hat{\Pi}_{ty}\right)\Psi_{t}^v + e^{-i\frac{\theta}{2}}_{c}\hat{U}_{0}^c\Psi_{b}^c &= E\Psi_{t}^c,\\
     v_F\left(\hat{\Pi}_{tx} + i\hat{\Pi}_{ty}\right)\Psi_{t}^c + \left(-\Delta + \hat{V}_t^v + \hbar\omega_c\hat{a}^{\dagger}\hat{a} + \hat{D}\right)\Psi_{t}^v + e^{i\frac{\theta}{2}}\hat{U}_{0}^v\Psi_{b}^v &=  E\Psi_{t}^v,\\ 
     \left(\Delta + \hat{V}_b^c + \hbar\omega_c\hat{a}^{\dagger}\hat{a} - \hat{D}\right)\Psi_{b}^c + v_F\left(\hat{\Pi}_{bx} - i\hat{\Pi}_{by}\right)\Psi_{b}^v + e^{i\frac{\theta}{2}}_{c}(\hat{U}^c_{0})^{\dagger}\Psi_{t}^c &= E\Psi_{b}^c,\\
     v_F\left(\hat{\Pi}_{bx} + i\hat{\Pi}_{by}\right)\Psi_{b}^c + \left(-\Delta + \hat{V}_b^v + \hbar\omega_c\hat{a}^{\dagger}\hat{a} - \hat{u}\right)\Psi_{b}^v + e^{-i\frac{\theta}{2}}(\hat{U}^v_{0})^{\dagger}\Psi_{t}^v &=  E\Psi_{b}^v,      
 \end{aligned}
 \end{equation}
 \end{widetext}
where for brevity we introduce the notation $\hat{\boldsymbol{\Pi}}_{b,t} = \hat{\mathbf{p}} - \hbar\boldsymbol{\kappa}_{+,-} + e\mathbf{A}^{(xy)}(\hat{a}+\hat{a}^{\dagger})$ and $\hat{D} = i\frac{\omega_c e A^{(z)} d}{2}\left(\hat{a} - \hat{a}^{\dagger}\right)$.

We consider a TMD material that has a large gap between the conduction and valence bands, and for states localized in the latter, we have $E \approx -\Delta$ and $||\Psi_{t}^c||, ||\Psi_{b}^c|| \ll ||\Psi_{t}^v||, ||\Psi_{b}^v|| $. Additionally, the moir\'{e} potentials and photon energy are also weak such that $||\hat{U}_0||, ||\hat{V}_t||, ||\hat{V}_b||, \hbar\omega_c \ll \Delta$. By neglecting higher order terms, the first two equations of (\ref{SM:eq-all 4}) corresponding to the top layer become:
\begin{widetext}
\begin{equation}
\begin{aligned}
    &2\Delta\Psi_{t}^c + v_F\left(\hat{\Pi}_{tx} - i\hat{\Pi}_{ty}\right)\Psi_{t}^v = 0,\\
    &v_F\left(\hat{\Pi}_{tx} + i\hat{\Pi}_{ty}\right)\Psi_{t}^c + \left(-\Delta + \hat{V}_t^v + \hbar\omega_c\hat{a}^{\dagger}\hat{a} - \hat{D}\right)\Psi_{t}^v + e^{i\frac{\theta}{2}}\hat{U}_{0}^v\Psi_{b}^v =  E\Psi_{t}^v.
\end{aligned}
\end{equation}
Solving the first line for $\Psi_{t}^c$ and defining the effective mass $m^{\star} = \frac{\Delta}{v_F^2}$, the second line can be written as
\begin{equation}
\left(-\frac{\left[\hat{\mathbf{p}} -\hbar\boldsymbol{\kappa}_-+ e\mathbf{A}^{(xy)}(\hat{a}+\hat{a}^{\dagger})\right]^2}{2m^{\star}} + \hat{V}_t^v - \Delta + \hbar\omega_c\hat{a}^{\dagger}\hat{a} + \hat{D}\right)\Psi_{t}^v + e^{i\frac{\theta}{2}}\hat{U}_{0}^v\Psi_{b}^v = E\Psi_{t}^v.
\end{equation}
Following the same procedure, we obtain another equation for the bottom layer with $\boldsymbol{\kappa}_- \to \boldsymbol{\kappa}_+$, $t \to b$. Combining the two equations we get an effective Hamiltonian for the valence bands: 
 \begin{equation}
\widetilde{\mathcal{H}} = 
    -\frac{1}{2m^{\star}}\begin{pmatrix}
   (\hat{\mathbf{p}} - \hbar\mathbf{\boldsymbol{\kappa}_-}  + e\mathbf{A}^{(xy)}(\hat{a}+\hat{a}^{\dagger}))^2 & 0 \\
    0 & (\hat{\mathbf{p}} - \hbar\boldsymbol{\kappa}_+ +e\mathbf{A}^{(xy)}(\hat{a}+\hat{a}^{\dagger}))^2
\end{pmatrix} 
    +  
    \begin{pmatrix}
    \hat{V}_t^v & \hat{U}_{0}^v \\ \hat{U}_{0}^{v\dagger} & \hat{V}_b^v
    \end{pmatrix}  + \hbar \omega_c \hat{a}^{\dagger}\hat{a} + i \omega_c\frac{eA^{(z)}d}{2}(\hat{a}-\hat{a}^{\dagger})\tau_z.
    \label{SM:eq-last H}
\end{equation}
 \end{widetext}
 The intralayer moir\'{e} potentials $\hat{V}^{v}_{t}$ and $\hat{V}^{v}_{b}$ can be written as \cite{Wu2019,Zhai2020}:
\begin{equation}
\label{eq:S-Vt}
    V_{t/b}^v(\mathbf{r}) = V_0 \sum_{i=1}^3\text{cos}(\mathbf{K}_i\cdot\boldsymbol{\Delta}(\mathbf{r}) \pm \alpha),
\end{equation}
where $\mathbf{K}_i$ are shown in Fig. \ref{fig:SM-TMD-mBZ}. In order to simplify the calculation, one could use the relation $\mathbf{K}_i\cdot\boldsymbol{\Delta
}(\mathbf{r}) = \mathbf{G}_i\cdot\mathbf{r}$, where $\mathbf{G}_{1,2}$ are reciprocal lattice vectors of the moir\'{e} system defined above and $\mathbf{G}_3 = \mathbf{G}_1 - \mathbf{G}_2$. The interlayer coupling $\hat{U}_0^v$ is given by \cite{Wu2019,Zhai2020}:
\begin{equation}
\label{eq:S:Uvv}
    U_{0}^v(\mathbf{r}) =
   h_0\sum_{j=1}^{3}  e^{i\left(\mathbf{G}_j-\mathbf{G}_1\right)\cdot\mathbf{r}}.
\end{equation}
For the numerical simulations presented in the paper we used the parameters of $\text{MoTe}_2$: $m^{\star} \approx 0.62m_e$ where $m_e$ is the bare electron mass, $a_0 \approx 3.5$\AA, $d \approx 10$\AA, moir\'{e} potentials $(V_0, \alpha, h_0) \approx (16\text{ meV}, 89.6^{\circ}, -8.5\text{ meV})$ \cite{Wu2019}.

\section{Particle-hole transformation}
In this section we detail the present model (\ref{SM:eq-last H}) in the hole picture. In second quantization, (\ref{SM:eq-last H}) reads:
\begin{widetext}
    \begin{equation}
        \begin{aligned}
            \widetilde{\mathcal{H}} &= \hbar\omega_c\hat{a}^{\dagger}\hat{a}-\frac{\hbar^2}{2m^{\star}}\sum_{\mathbf{q},\nu}\hat{c}_{\nu\mathbf{q}}^{\dagger}\left[\mathbf{q} -\mathbf{K}_{\nu}+ \frac{e\mathbf{A}^{(xy)}}{\hbar}(\hat{a}+\hat{a}^{\dagger})\right]^2\hat{c}_{\nu\mathbf{q}} + \sum_{\nu}\int d\mathbf{r}\hat{c}_{\nu}^{\dagger}(\mathbf{r})V^v_{\nu}(\mathbf{r})\hat{c}_{\nu}(\mathbf{r})\\
            &+ \int d\mathbf{r}\left[\hat{c}_{t}^{\dagger}(\mathbf{r})U_0^v(\mathbf{r})\hat{c}_{b}(\mathbf{r}) + \text{h.c}\right] + i\omega_c\frac{eA^{(z)}d}{2}\left(\hat{a}-\hat{a}^{\dagger}\right)\int d\mathbf{r}\left(\hat{c}^{\dagger}_t(\mathbf{r})\hat{c}_{t}(\mathbf{r}) - \hat{c}^{\dagger}_b\mathbf{r}\hat{c}_{b}\right),
        \end{aligned}
    \end{equation}
where $\nu = t,b$ and $\mathbf{K}_{b,t} = \boldsymbol{\kappa}_{+,-}$. The particle-hole transformation can be defined by a unitary operator $\hat{\Gamma}$, such that $\hat{\Gamma} \hat{c}^{\dagger}_{\nu}(\mathbf{r}) \hat{\Gamma}^{\dagger} = \hat{c}_{\nu}(\mathbf{r})$. Under this transformation the momentum operators transform as $\hat{c}_{\nu\mathbf{q}}^{\dagger} \rightarrow \hat{\Gamma} \hat{c}^{\dagger}_{\nu\mathbf{q}} \hat{\Gamma}^{\dagger} = \hat{c}_{\nu-\mathbf{q}}$. Using the particle-hole transformation, the Hamiltonian $\hat{\mathcal{H}} = \hat{\Gamma}\widetilde{H}\hat{\Gamma}^{\dagger}$ becomes: 
    \begin{equation}
    \label{SM:eq-2hole}
        \begin{aligned}
            \hat{\mathcal{H}} &= \hbar\omega_c\hat{a}^{\dagger}\hat{a}+\frac{\hbar^2}{2m^{\star}}\sum_{\mathbf{q},\nu}\hat{c}_{\nu\mathbf{q}}^{\dagger}\left[\mathbf{q} +\mathbf{K}_{\nu}- \frac{e\mathbf{A}^{(xy)}}{\hbar}(\hat{a}+\hat{a}^{\dagger})\right]^2\hat{c}_{\nu\mathbf{q}} - \sum_{\nu}\int d\mathbf{r}\hat{c}_{\nu}^{\dagger}(\mathbf{r})V^v_{\nu}(\mathbf{r})\hat{c}_{\nu}(\mathbf{r})\\
            &- \int d\mathbf{r}\left[\hat{c}_{b}^{\dagger}(\mathbf{r})U_0^v(\mathbf{r})\hat{c}_{t}(\mathbf{r}) + \text{h.c}\right] - i\omega_c\frac{eA^{(z)}d}{2}\left(\hat{a}-\hat{a}^{\dagger}\right)\int d\mathbf{r}\left(\hat{c}^{\dagger}_t(\mathbf{r})\hat{c}_{t}(\mathbf{r}) - \hat{c}^{\dagger}_b(\mathbf{r})\hat{c}_{b}(\mathbf{r})\right).
        \end{aligned}
    \end{equation}
\end{widetext}
Equation (\ref{SM:eq-2hole}) results in the change of the sign of the effective mass, the charge and the moir\'{e} potentials. Note that we have disregarded divergent sums that result from the interaction of all valance electrons with the photonic field. This is justified if we assume the light-matter interaction vanishes when all valence mini-bands are completely filled.
\section{Purity of mixed states}
We provide here more details about the purity of the electronic reduced density matrix that is discussed in the manuscript.  By diagonalizing the light-matter Hamiltonian in Equation (3) in the main text, one obtains electron-photon eigenstates $\vert \Psi^{{\mathrm{(e-p})}}\rangle$ that belong to the Hilbert space, which is the product of the electron and photon Hilbert spaces. If one wants to focus on the electronic properties only, the standard approach is to consider the reduced density matrix defined as \cite{CohenTannoudji2020} $\hat{\rho} = \text{Tr}_{\text{phot}}\left(\vert \Psi^{{\mathrm{(e-p})}}\rangle \langle \Psi^{{\mathrm{(e-p})}}\vert\right)$, where the trace is here over the photonic degrees of freedom. 
The purity of a density matrix is defined as $\mathbb{P} = \text{Tr}_{\text{el}}\left(\hat{\rho}^2\right)$, where here the trace is over the electronic degrees of freedom. Indeed, when the reduced density matrix describes a pure state, the purity is equal to 1, since $\hat{\rho}^2 = \hat{\rho} $ in that case. This occurs for electron-photon states that are separable, namely of the form $\vert \Psi^{{\mathrm{(e-p})}}\rangle = \vert \psi^{(e)}\rangle\otimes\vert \phi^{(p)}\rangle$, implying that the purity is merely $\text{Tr}_{\text{el}}\left(\hat{\rho}^2\right) = \text{Tr}_{\text{el}}\left(\hat{\rho}\right) = \text{Tr}_{\text{el}}\left(\vert \psi^{{\mathrm{(e})}}\rangle \langle \psi^{{\mathrm{(e})}}\vert\right) = 1$. Instead, when the reduced density matrix is mixed, it can be written as $\hat{\rho} = \sum_{\lambda = 1}^{d}p_{\lambda}\vert \psi^{(e)}_{\lambda}\rangle \langle \psi^{(e)}_{\lambda}\vert$ where $\{\vert \psi^{(e)}_{\lambda}\rangle\}$ is an  orthonormal basis of an electronic Hilbert subspace of dimension $d$ and $p_{\lambda}$ is the probability of finding the state in $\vert\psi^{(e)}_{\lambda}\rangle$, with $0 \leq p_{\lambda} \leq 1$ and $\sum_{\lambda=1}^{d}p_{\lambda} = 1$. 
These conditions lead to $\text{Tr}_{\text{el}}\left(\hat{\rho}^2\right) = \sum_{\lambda}p^2_{\lambda} \leq \sum_{\lambda}p_{\lambda} = 1$. Moreover, the Cauchy–Schwarz inequality gives $\sum_{\lambda}p^2_{\lambda} \geq 1/d$: hence, the purity reaches its minimum value $1/d$ if and only if $\forall \lambda, p_{\lambda} = 1/d$, which is known as a maximally mixed state. When the dimension $d$ tends to infinity, the purity tends to $0$. 


\section{Effective electronic Hamiltonian in high purity regime}
In this section, we provide details about the effective electronic Hamiltonian obtained by adiabatic elimination of the photonic degree of freedom \cite{Ciuti2021,Arwas2023}. Moreover, we use this theory to compare to the topological diagram in Fig. 2 of the main text, where the electron purity is very close to 1. Fig. 2 in the manuscript has been calculated using the exact electron-photon eigenstates and the electron-photon Chern number. The topological phase boundary in Fig. 2 can be calculated approximately  by using the eigenstates of the effective electronic Hamiltonian and the traditional electron Chern number. Such approach works well in the regime considered in Fig. 2, where the electronic purity is close to 1.

Let us see here the details of the effective electronic Hamiltonian.
Eq. (3) in main text can be written as $\hat{\mathcal{H}} = \hat{\mathcal{H}}_0 + \hbar\omega_c\hat{a}^{\dagger}\hat{a} + D(\hat{a}+\hat{a}^{\dagger})^2 + \hat{V}_{xy}(\hat{a}+\hat{a}^{\dagger}) + \hat{V}_zi(\hat{a}-\hat{a}^{\dagger})$, where $\hat{\mathcal{H}}_0$ is the bare electronic Hamiltonian with its eigenstates $\vert\psi^0_{n\mathbf{k}}\rangle$ and energies $\epsilon^{0}_{n\mathbf{k}}$. The term $D = e^2|\mathbf{A}^{(xy)}|^2/(2m^{\star})$ is due  to the diamagnetic energy, while $\hat{V}_{xy}$ and $\hat{V}_z$ are respectively in-plane and out-of-plane coupling term with the following form:
\begin{equation}
    \begin{aligned}
        &\hat{V}_{xy} = -\frac{e}{m^{\star}}\mathbf{A}^{(xy)}\cdot\begin{pmatrix}
            \hat{\mathbf{p}} + \hbar\boldsymbol{\kappa}_- && 0 \\ 0 && \hat{\mathbf{p}} + \hbar\boldsymbol{\kappa}_-
        \end{pmatrix},\\
        &\hat{V}_z = \frac{\omega_c e A^{(z)} d}{2}\begin{pmatrix}
            1 && 0 \\ 0 && -1
        \end{pmatrix}.
    \end{aligned}
\end{equation}
By performing a Bogoliubov transformation for the photonic degrees of freedom, we can diagonalize the photonic part in the form $\hbar\omega_c\hat{a}^{\dagger}\hat{a} + D(\hat{a}+\hat{a}^{\dagger})^2 = \hbar\tilde{\omega}_c\hat{\alpha}^{\dagger}\hat{\alpha}$, where the dressed photon frequency  reads $\tilde{\omega}_c = \sqrt{\omega^2_c + 2\omega_ce^2|\mathbf{A}^{(xy)}|^2/(m^{\star}\hbar)}$ and the corresponding dressed boson operators are such that: $\hat{a} + \hat{a}^{\dagger} = \sqrt{\omega_c/\tilde{\omega}_c}(\hat{\alpha}+\hat{\alpha}^{\dagger})$ and $\hat{a} - \hat{a}^{\dagger} = \sqrt{\tilde{\omega}_c/\omega_c}(\hat{\alpha}-\hat{\alpha}^{\dagger})$. At the lowest order in perturbation theory, there is an effective coupling between different miniband states via an an intermediate state with a (dressed) cavity photon. Namely, we have 
\begin{equation}
 \vert\psi_{n\mathbf{k}}^0\rangle\vert 0\rangle \rightarrow \vert\psi_{\lambda\mathbf{k}}^0\rangle\vert 1\rangle \rightarrow \vert\psi_{n'\mathbf{k}}^0\rangle\vert 0\rangle.
\end{equation}
Such a process couples $\vert\psi^0_{n\mathbf{k}}\rangle$ and $\vert\psi^0_{n'\mathbf{k}}\rangle$ with a corresponding coupling energy:
\begin{equation}
    \Gamma_{n'n\mathbf{k}} \simeq \frac{1}{2}\sum_{\lambda\mathbf{k}}\left(\frac{V_{n'\lambda\mathbf{k}}V_{\lambda n\mathbf{k}}}{\epsilon^0_{\lambda\mathbf{k}}+\hbar\tilde{\omega}_c-\epsilon^0_{n'\mathbf{k}}} + \frac{V_{n'\lambda\mathbf{k}}V_{\lambda n\mathbf{k}}}{\epsilon^0_{\lambda\mathbf{k}}+\hbar\tilde{\omega}_c -\epsilon^0_{n\mathbf{k}}}\right),
\label{eq:SM:coeff}
\end{equation}
with $V_{n'n\mathbf{k}} = \langle \psi_{n'\mathbf{k}}^0\vert \beta \hat{V}_{xy} + \beta^{-1}\hat{V}_z \vert \psi_{n\mathbf{k}}^0 \rangle$ and $\beta = \sqrt{\omega_c/\tilde{\omega}_c}$. The effective electronic Hamiltonian reads:
\begin{equation}
    \hat{\mathcal{H}}_{\text{eff}} = \sum_{n,\mathbf{k}}\epsilon^0_{n\mathbf{k}}\vert\psi_{n\mathbf{k}}^0\rangle \langle\psi_{n\mathbf{k}}^0\vert + \sum_{n',n,\mathbf{k}}\Gamma_{n'n\mathbf{k}}\vert\psi_{n'\mathbf{k}}^0\rangle \langle\psi_{n\mathbf{k}}^0\vert.
\label{eq:SM:Heff}
\end{equation}
Equations (\ref{eq:SM:coeff}) and (\ref{eq:SM:Heff}) are valid only when the photon can be adiabatically eliminated. The agreement in Fig. 2 of the main text is excellent with only small variations at larger couplings due to the small deviations of the electron purity from unity with the parameters considered there.   
    
\section{Electronic spectral function and spatial distribution}
\begin{figure}[t!]
\hspace{5pt}\includegraphics[width = 0.95\hsize]{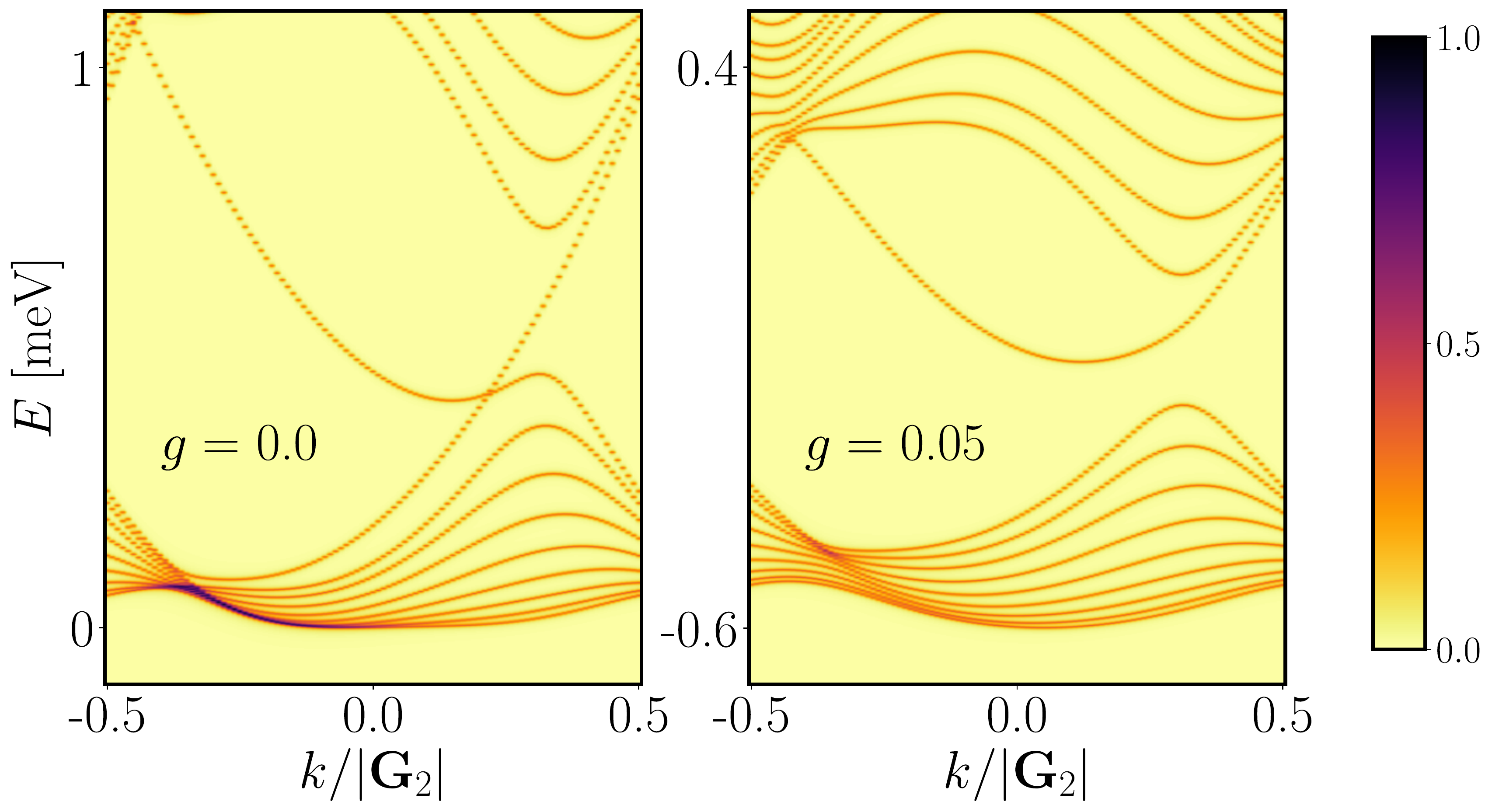}
\includegraphics[width = 0.95\hsize]{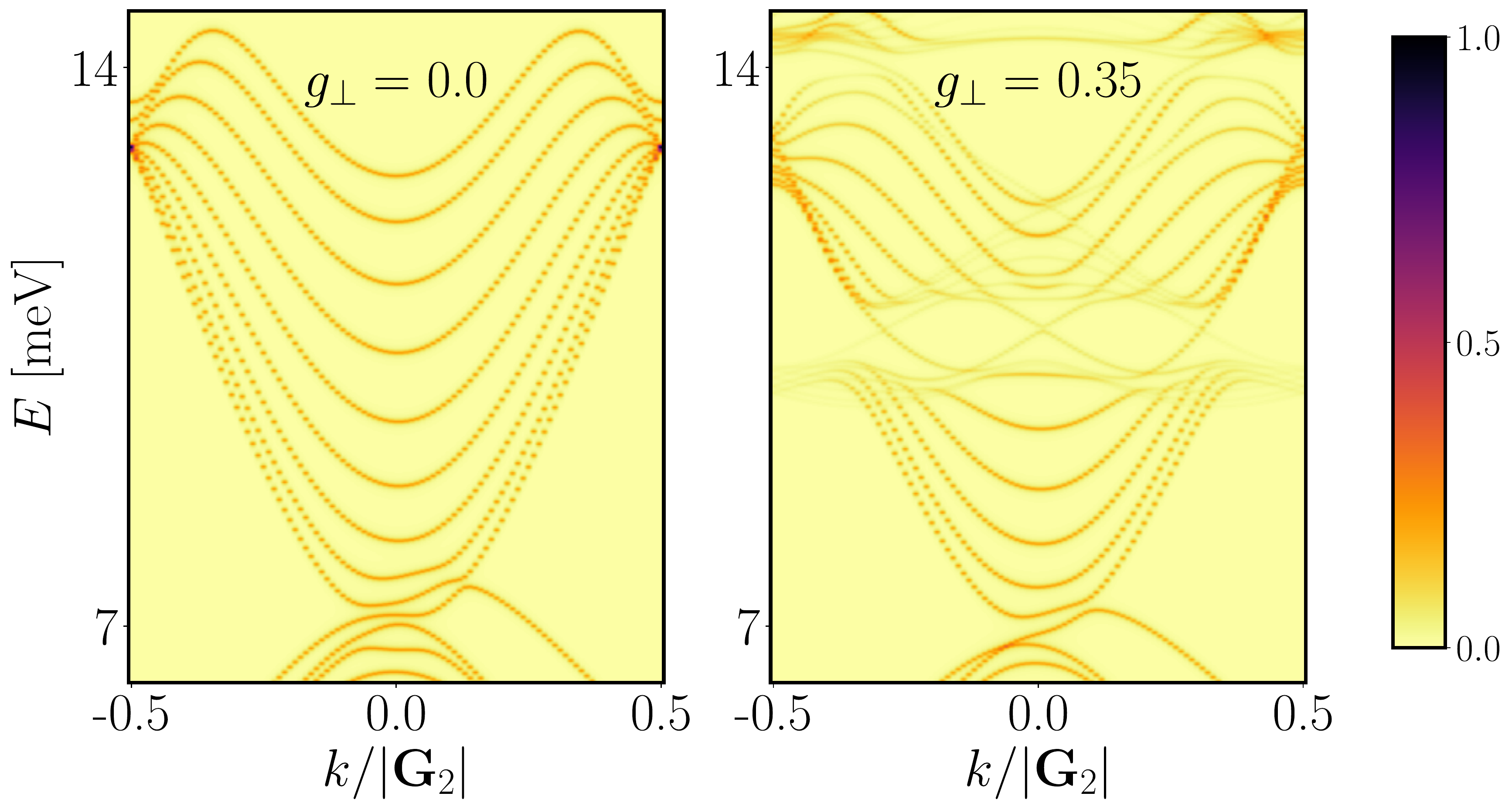}
    \caption{Color plot of the electron spectral function  (amplitude normalized by its maximum value) corresponding to the off-resonant (top panels) and resonant (bottom panels) configurations considered in Fig. 3 and 5 of the manuscript, respectively.}
  \label{SM-Fig-SF}
\end{figure}

\begin{figure}[b!]
\includegraphics[width = 0.45\hsize]{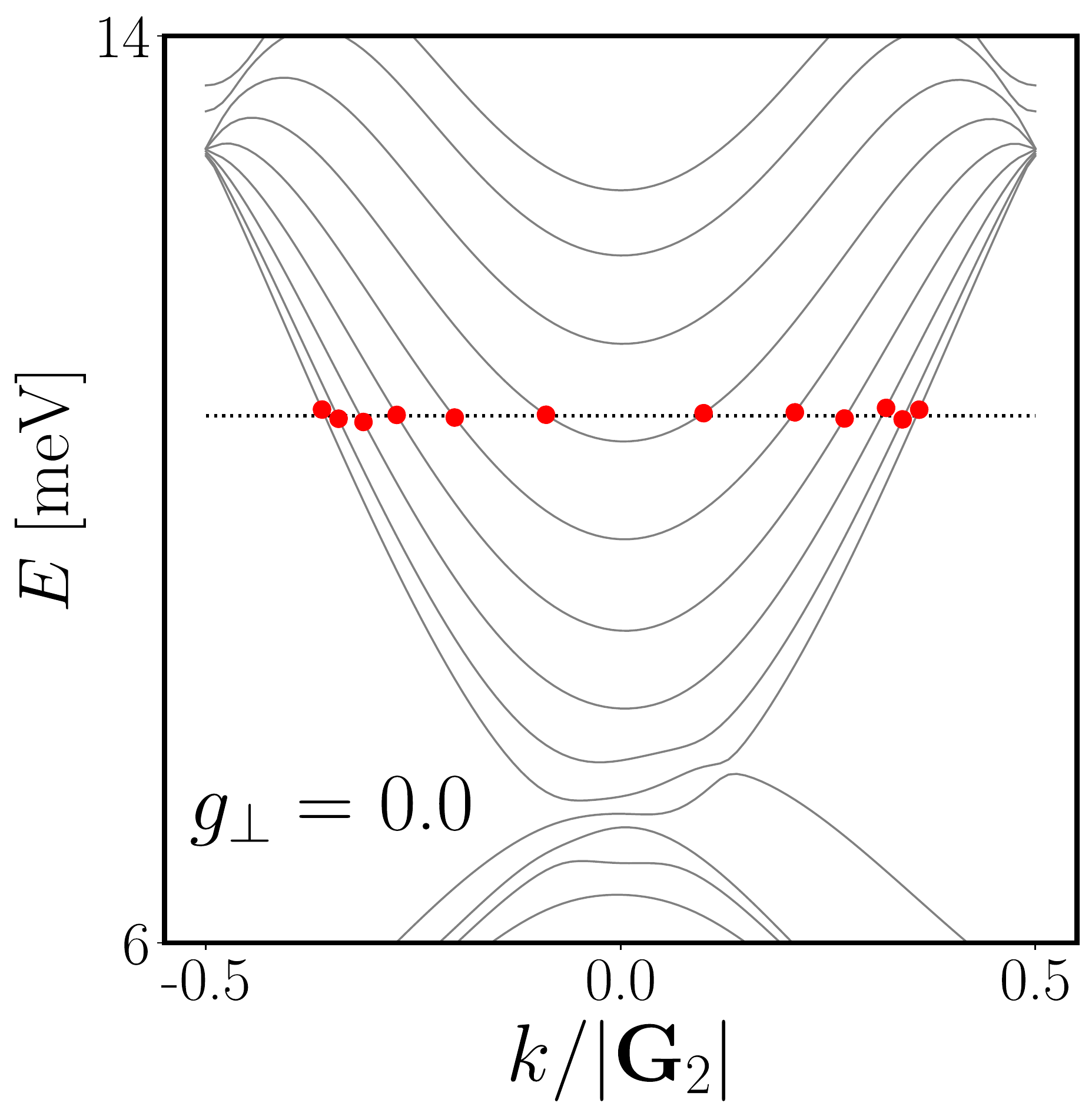}
\includegraphics[width = 0.49\hsize]{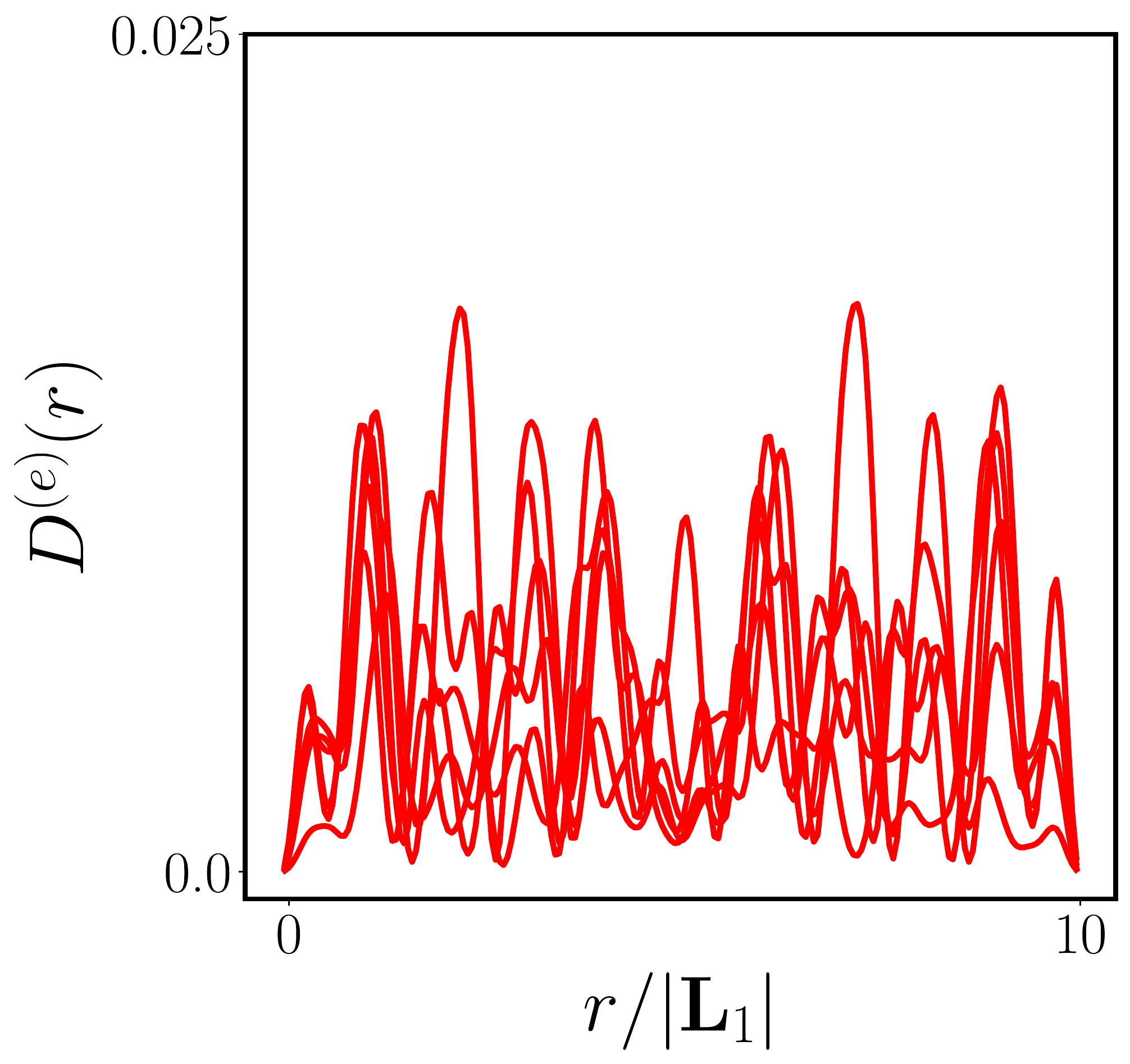}\\
\includegraphics[width = 0.45\hsize]{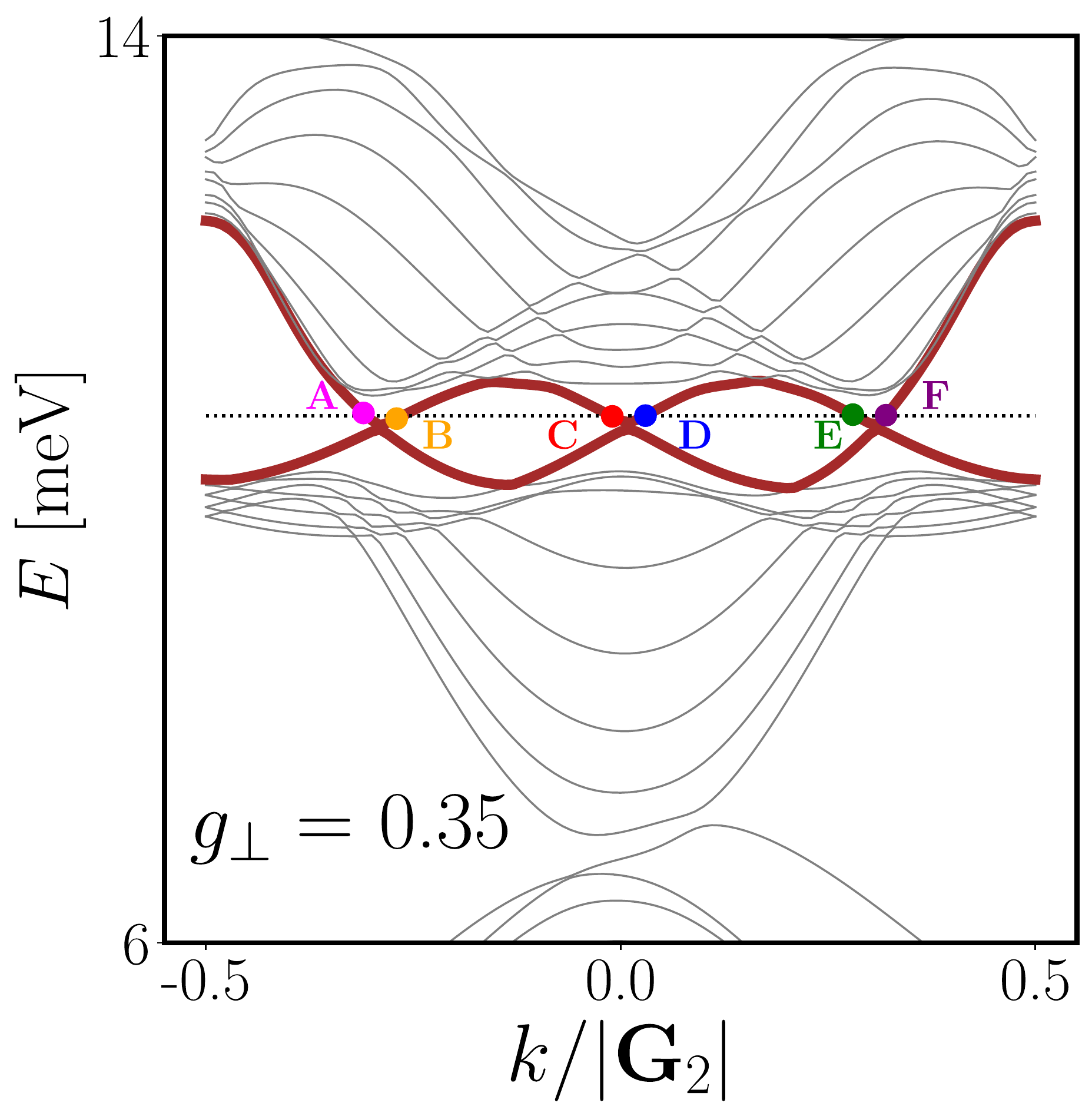}
\includegraphics[width = 0.49\hsize]{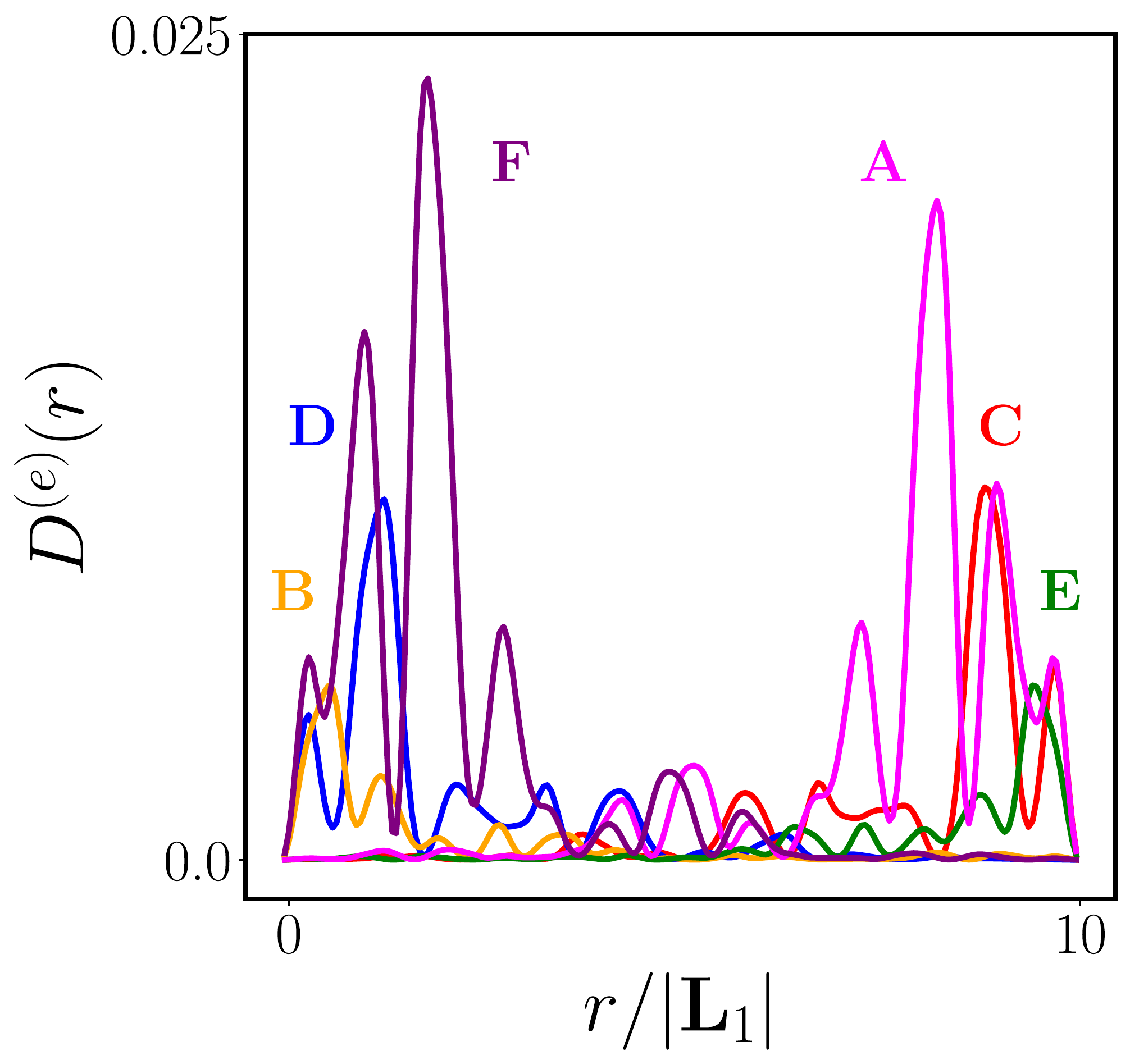}
\caption{Spatial distribution of the states indicated in the energy-momentum diagram for the electron-photon eigenstates. The top panels correspond to no cavity coupling, while the botton panels correspond to $g_{\perp} = 0.35$. These plots correspond to the same configuration (perpendicular polarization) and parameters as in Fig. 5 of the manuscript.}
\label{fig:SM-distribution}
\end{figure}

In this section, we provide some details about the electron spectral function, a quantity that can be measured via Angular Resolved Photo-Emission Spectroscopy (ARPES) and the spatial distribution of the edge states, which can be measured for example via Scanning Tunneling Microscopy (STM) or Microwave Impedance Microscopy.
In this section, we will consider ribbon samples with open boundary conditions in the short direction.
The spatially resolved electronic density of state is given by:
\begin{widetext}
\begin{equation}
    A(\mathbf{k},\omega,r_1) = \sum_n\sum_{\nu = t,b}\int dr_2\left|\langle\Psi^{(e-p)}_{n\mathbf{k}}\vert r_1,r_2,\nu,\mathbf{k},0\rangle\right|^2\delta\left(\omega - \mathcal{E}^{(e-p)}_{n\mathbf{k}}\right),
\label{SM:eq-SF}
\end{equation}
\end{widetext}
where $r_1$ ($r_2$) corresponds to the position in the short (long) direction, $\nu = t,b$ describes the layer degree of freedom. Moreover, $\vert\Psi^{(e-p)}_{n\mathbf{k}}\rangle$ and $\mathcal{E}^{(e-p)}_{n\mathbf{k}}$ are respectively electron-photon eigenstates and energies of the Hamiltonian, while $\vert0\rangle$ is a zero-photon state. Note that the domain of $r_2$ is restricted to one unit cell. From (\ref{SM:eq-SF}), the spectral function is directly obtained by taking the integral $\int dr_1$ of the spectral function $A$. Detailed plots of this
observable are reported in Fig. \ref{SM-Fig-SF}, which has the same parameters as in Fig. 3 and 5 of the manuscript. 
\\
Given an electron-photon eigenstate $\vert \Psi^{(e-p)}_{n\mathbf{k}}\rangle$, its spatial distribution on the short direction (open boundary) is calculated by tracing over all the other degrees of freedom, namely: 
\begin{equation}
    \mathcal{D}^{(e-p)}_{n\mathbf{k}}(r_1) = \text{Tr}_{\text{phot},r_2,\nu}\left(\vert\Psi^{(e-p)}_{n\mathbf{k}}\rangle\langle\Psi^{(e-p)}_{n\mathbf{k}}\vert\right).
    \label{SM:eq-ep distri}
\end{equation}
However, equation (\ref{SM:eq-ep distri}) is not the electronic local density of states. Instead from (\ref{SM:eq-SF}) the electronic local density of states is calculated as:
\begin{equation}
\label{SM:eq-e distri}  
\begin{aligned}
    D^{(e)}_{n\mathbf{k}}(r_1) &= \int_{\mathcal{E}^{(e-p)}_{n\mathbf{k}}-\Delta\mathcal{E}}^{\mathcal{E}^{(e-p)}_{n\mathbf{k}}+\Delta\mathcal{E}} d\omega A(\mathbf{k},\omega,r_1)\\
    &\simeq\sum_{\nu = t,b}\int dr_2\left|\langle\Psi^{(e-p)}_{n\mathbf{k}}\vert r_1,r_2,\nu,\mathbf{k},0\rangle\right|^2\, .
\end{aligned} 
\end{equation}
The spatial distribution of the edge states created by the cavity coupling is reported in Fig. \ref{fig:SM-distribution}.

\bibliography{Bib.bib}